\pdfoutput=1

\documentclass[12pt,a4paper]{article}

\usepackage{ifthen} 
\newboolean{pdflatex}
\setboolean{pdflatex}{true} 

\newboolean{articletitles}
\setboolean{articletitles}{true} 

\newboolean{uprightparticles}
\setboolean{uprightparticles}{false} 

\newboolean{inbibliography}
\setboolean{inbibliography}{false} 

\usepackage{lineno}  
\usepackage{xspace} 
\usepackage{caption} 

\usepackage{graphicx}  
\usepackage{color}
\usepackage{colortbl}
\graphicspath{{./figs/}} 

\usepackage{booktabs}
\usepackage{amsmath} 
\usepackage{amssymb}
\usepackage{amsfonts}
\usepackage{upgreek} 

\newcommand*\patchAmsMathEnvironmentForLineno[1]{%
\expandafter\let\csname old#1\expandafter\endcsname\csname #1\endcsname
\expandafter\let\csname oldend#1\expandafter\endcsname\csname
end#1\endcsname
 \renewenvironment{#1}%
   {\linenomath\csname old#1\endcsname}%
   {\csname oldend#1\endcsname\endlinenomath}%
}
\newcommand*\patchBothAmsMathEnvironmentsForLineno[1]{%
  \patchAmsMathEnvironmentForLineno{#1}%
  \patchAmsMathEnvironmentForLineno{#1*}%
}
\AtBeginDocument{%
\patchBothAmsMathEnvironmentsForLineno{equation}%
\patchBothAmsMathEnvironmentsForLineno{align}%
\patchBothAmsMathEnvironmentsForLineno{flalign}%
\patchBothAmsMathEnvironmentsForLineno{alignat}%
\patchBothAmsMathEnvironmentsForLineno{gather}%
\patchBothAmsMathEnvironmentsForLineno{multline}%
\patchBothAmsMathEnvironmentsForLineno{eqnarray}%
}

\usepackage{hyperref}    
\usepackage[all]{hypcap} 

\usepackage{xspace} 
\usepackage{upgreek}

\def\lhcb {\mbox{LHCb}\xspace}

\def\herschel {\mbox{\textsc{HeRSCheL}}\xspace}

\def\MagUp {\mbox{\em Mag\kern -0.05em Up}\xspace}

\ifthenelse{\boolean{uprightparticles}}%
{

 \def\Ppsi        {\ensuremath{\uppsi}\xspace}

 \def\PDelta      {\ensuremath{\Delta}\xspace}                 
 \def\PXi      {\ensuremath{\Xi}\xspace}                 
 \def\PLambda      {\ensuremath{\Lambda}\xspace}                 
 \def\PSigma      {\ensuremath{\Sigma}\xspace}                 
 \def\POmega      {\ensuremath{\Omega}\xspace}                 
 \def\PUpsilon      {\ensuremath{\Upsilon}\xspace}                 
 
 \def\PB      {\ensuremath{\mathrm{B}}\xspace}                 
                  
 \def\PD      {\ensuremath{\mathrm{D}}\xspace}

 \def\PJ      {\ensuremath{\mathrm{J}}\xspace}                 
 \def\PK      {\ensuremath{\mathrm{K}}\xspace}

 \def\Pi      {\ensuremath{\mathrm{i}}\xspace}

}
{

 \def\Ppsi        {\ensuremath{\psi}\xspace}                 
                  
 \mathchardef\PDelta="7101
 \mathchardef\PXi="7104
 \mathchardef\PLambda="7103
 \mathchardef\PSigma="7106
 \mathchardef\POmega="710A
 \mathchardef\PUpsilon="7107
                  
 \def\PB      {\ensuremath{B}\xspace}                 
                  
 \def\PD      {\ensuremath{D}\xspace}

 \def\PJ      {\ensuremath{J}\xspace}                 
 \def\PK      {\ensuremath{K}\xspace}

 \def\Pi      {\ensuremath{i}\xspace}

}

\makeatletter
\ifcase \@ptsize \relax
  \newcommand{\miniscule}{\@setfontsize\miniscule{4}{5}}
\or
  \newcommand{\miniscule}{\@setfontsize\miniscule{5}{6}}
\or
  \newcommand{\miniscule}{\@setfontsize\miniscule{5}{6}}
\fi
\makeatother

\DeclareRobustCommand{\optbar}[1]{\shortstack{{\miniscule (\rule[.5ex]{1.25em}{.18mm})}
  \\ [-.7ex] $#1$}}

  \def\Kbar    {{\kern 0.2em\overline{\kern -0.2em \PK}{}}\xspace}

\def\KorKbar    {\kern 0.18em\optbar{\kern -0.18em K}{}\xspace}

  \def\Dbar    {{\kern 0.2em\overline{\kern -0.2em \PD}{}}\xspace}

\def\DorDbar    {\kern 0.18em\optbar{\kern -0.18em D}{}\xspace}

\def\Bbar    {{\ensuremath{\kern 0.18em\overline{\kern -0.18em \PB}{}}}\xspace}

\def\BorBbar    {\kern 0.18em\optbar{\kern -0.18em B}{}\xspace}

\def\jpsi     {{\ensuremath{{\PJ\mskip -3mu/\mskip -2mu\Ppsi\mskip 2mu}}}\xspace}

  \def\Y#1S{\ensuremath{\PUpsilon{(#1S)}}\xspace}

\def\Lbar        {{\ensuremath{\kern 0.1em\overline{\kern -0.1em\PLambda}}}\xspace}
\def\LorLbar    {\kern 0.18em\optbar{\kern -0.18em \PLambda}{}\xspace}

\def\AT#1     {\ensuremath{A_{\mathrm{T}}^{#1}}\xspace}           

\def\C#1      {\ensuremath{\mathcal{C}_{#1}}\xspace}                       
\def\Cp#1     {\ensuremath{\mathcal{C}_{#1}^{'}}\xspace}                    
\def\Ceff#1   {\ensuremath{\mathcal{C}_{#1}^{\mathrm{(eff)}}}\xspace}        
\def\Cpeff#1  {\ensuremath{\mathcal{C}_{#1}^{'\mathrm{(eff)}}}\xspace}       
\def\Ope#1    {\ensuremath{\mathcal{O}_{#1}}\xspace}                       
\def\Opep#1   {\ensuremath{\mathcal{O}_{#1}^{'}}\xspace}                    

\newcommand{\tev}{\ifthenelse{\boolean{inbibliography}}{\ensuremath{~T\kern -0.05em eV}\xspace}{\ensuremath{\mathrm{\,Te\kern -0.1em V}}}\xspace}
\newcommand{\gev}{\ensuremath{\mathrm{\,Ge\kern -0.1em V}}\xspace}
\newcommand{\mev}{\ensuremath{\mathrm{\,Me\kern -0.1em V}}\xspace}
\newcommand{\kev}{\ensuremath{\mathrm{\,ke\kern -0.1em V}}\xspace}
\newcommand{\ev}{\ensuremath{\mathrm{\,e\kern -0.1em V}}\xspace}
\newcommand{\gevc}{\ensuremath{{\mathrm{\,Ge\kern -0.1em V\!/}c}}\xspace}
\newcommand{\mevc}{\ensuremath{{\mathrm{\,Me\kern -0.1em V\!/}c}}\xspace}
\newcommand{\gevcc}{\ensuremath{{\mathrm{\,Ge\kern -0.1em V\!/}c^2}}\xspace}
\newcommand{\gevgevcc}{\ensuremath{{\mathrm{\,Ge\kern -0.1em V^2\!/}c^2}}\xspace}
\newcommand{\gevgevcccc}{\ensuremath{{\mathrm{\,Ge\kern -0.1em V^2\!/}c^4}}\xspace}
\newcommand{\mevcc}{\ensuremath{{\mathrm{\,Me\kern -0.1em V\!/}c^2}}\xspace}

\def\ns   {\ensuremath{{\mathrm{ \,ns}}}\xspace}

\def\gsim{{~\raise.15em\hbox{$>$}\kern-.85em
          \lower.35em\hbox{$\sim$}~}\xspace}
\def\lsim{{~\raise.15em\hbox{$<$}\kern-.85em
          \lower.35em\hbox{$\sim$}~}\xspace}

\def\pt         {\mbox{$p_{\mathrm{ T}}$}\xspace}
\def\ptsq     	{\mbox{$p^2_{\rm T}$}\xspace}

\def\gauss      {\mbox{\textsc{Gauss}}\xspace}
\def\geant      {\mbox{\textsc{Geant4}}\xspace}

\def\pythia     {\mbox{\textsc{Pythia}}\xspace}

\def\tell1  {TELL1\xspace}
\def\ukl1   {UKL1\xspace}

\def\kV   {\ensuremath{\mbox{\,kV}}\xspace}

\newcommand{\ie}{\mbox{\itshape i.e.}\xspace}

\usepackage{cite} 
\usepackage{mciteplus}

\usepackage{longtable} 
\usepackage{booktabs}
\begin{document}

\renewcommand{\thefootnote}{\fnsymbol{footnote}}
\setcounter{footnote}{1}


\begin{titlepage}
\pagenumbering{roman}

\vspace*{-1.5cm}
\centerline{\large EUROPEAN ORGANIZATION FOR NUCLEAR RESEARCH (CERN)}
\vspace*{1.5cm}
\noindent
\begin{tabular*}{\linewidth}{lc@{\extracolsep{\fill}}r@{\extracolsep{0pt}}}
\ifthenelse{\boolean{pdflatex}}
{\vspace*{-2.7cm}\mbox{\!\!\!\includegraphics[width=.14\textwidth]{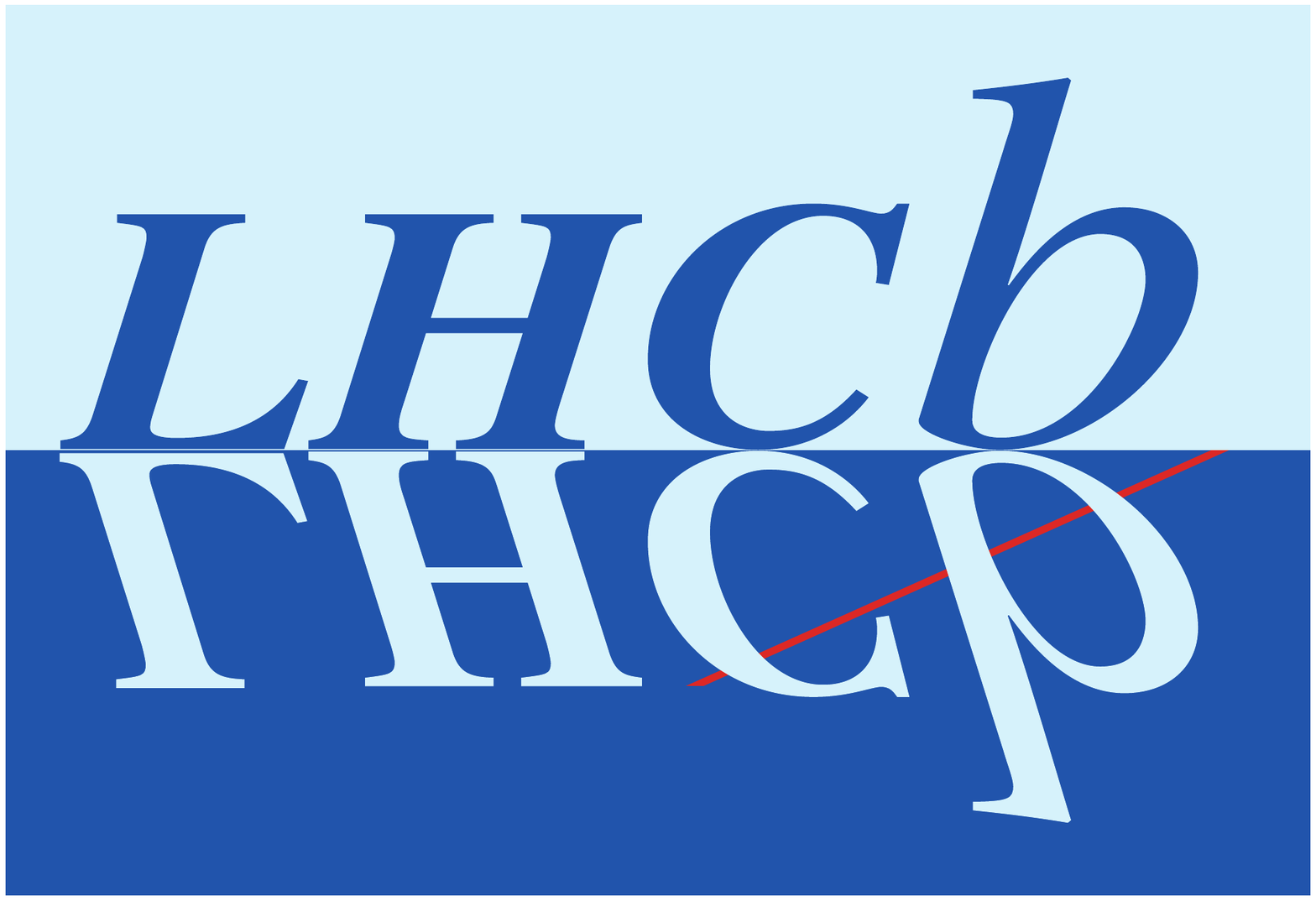}} & &}%
{\vspace*{-1.2cm}\mbox{\!\!\!\includegraphics[width=.12\textwidth]{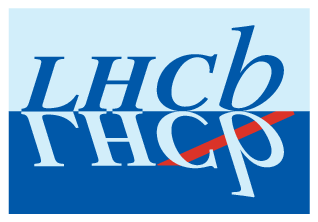}} & &}%
\\
 & & LHCb-DP-2016-003 \\  
 & & January 12, 2018 \\ 
\end{tabular*}

\vspace*{2.0cm}

{\normalfont\bfseries\boldmath\huge
\begin{center}
  The \herschel detector: high-rapidity shower counters for LHCb
\end{center}
}

\vspace*{2.0cm}

\begin{center}
K.~Carvalho Akiba$^1$,
F.~Alessio$^2$,
N.~Bondar$^{2,3}$,
W.~Byczynski$^2$,
V.~Coco$^2$,
P.~Collins$^2$,
R.~Dumps$^2$,
R.~Dzhelyadin$^4$,
P.~Gandini$^5$,
B.R.~Gruberg~Cazon$^5$,
R.~Jacobsson$^2$,
D.~Johnson$^2$,
J.~Manthey$^2$,
J.~Mauricio$^6$,
R.~McNulty$^7$,
S.~Monteil$^8$,
B.~Rachwal$^9$,
M.~Ravonel Salzgeber$^2$,
L.~Roy$^2$,
H.~Schindler$^2$,
S.~Stevenson$^5$,
G.~Wilkinson$^{5}$
\bigskip\\
{\it\footnotesize
$ ^1$ Universidade Federal do Rio de Janiero (UFRJ), Rio de Janeiro, Brazil\\
$ ^2$ European Organization for Nuclear Research (CERN), Geneva, Switzerland \\
$ ^3$ Petersburg Nuclear Physics Institute (PNPI), Gatchina, Russia\\
$ ^4$ Institute for High Energy Physics (IHEP), Protvino, Russia \\
$ ^5$ Department of Physics, University of Oxford, Oxford, United Kingdom \\
$ ^6$ FQA, ICC, Universitat de Barcelona, Avinguda Diagonal 647, Barcelona, Spain \\
$ ^7$ School of Physics, University College Dublin, Dublin, Ireland \\
$ ^8$ Clermont Universit\'e, Universit\'e Blaise Pascal, CNRS/IN2P3, LPC, Clermont-Ferrand, France \\
$ ^{9}$ AGH - University of Science and Technology, Faculty of Physics and Applied Computer Science, Krak{\'o}w, Poland
}
\end{center}

\vspace{\fill}

\begin{abstract}
  \noindent 
The \herschel detector consists of a set of scintillating counters, designed to increase the coverage of the \lhcb experiment in the high-rapidity regions on either side of the main spectrometer. The new detector improves the capabilities of \lhcb for studies of diffractive interactions, most notably Central Exclusive Production. In this paper the construction, installation, commissioning, and performance of \herschel are presented.
  
\end{abstract}


\begin{center}
  Submitted to JINST
\end{center}

\vspace{\fill}

{\footnotesize 
\centerline{\copyright~CERN on behalf of the \lhcb collaboration, licence \href{http://creativecommons.org/licenses/by/4.0/}{CC-BY-4.0}.}}
\vspace*{2mm}

\end{titlepage}


\newpage
\setcounter{page}{2}
\mbox{~}

\cleardoublepage

\newcommand{\red}[1]{\textcolor{red}{ #1}}
\renewcommand{\thefootnote}{\arabic{footnote}}
\setcounter{footnote}{0}



\pagestyle{plain} 
\setcounter{page}{1}
\pagenumbering{arabic}


%

\section{Introduction}
\label{sec:Introduction}

 \herschel (High Rapidity Shower Counters for LHCb) is a system of Forward Shower Counters (FSCs) located in the LHC tunnel on both sides of the LHCb interaction point.  
It was installed for Run 2 of the LHC, which began in 2015, with the aim of  enhancing LHCb's capabilities in diffractive physics, in particular Central Exclusive Production (CEP) analyses~\cite{Albrow:2010yb}.

Each FSC comprises a quadrant of scintillator planes equipped with PMTs, which are read out synchronously with the sub-detectors of the LHCb spectrometer.  The planes are situated close to the beampipe and detect showers induced by very forward particles interacting in the beampipe or other machine elements~\cite{Albrow:2014yma,Albrow:2014lta}. In this manner \herschel provides sensitivity to activity at higher rapidities than is available from the other sub-detectors of the experiment.  This increased acceptance will be valuable in the classification of different production processes in LHC collisions, for example the isolation of CEP candidates.

This paper is organised as follows.  The physics motivation for \herschel is elaborated in Sect.~\ref{sec:motivation}. The apparatus is described in Sect.~\ref{sec:setup}, and the calibration procedure and detector sensitivity are discussed in Sect.~\ref{sec:characterisation}. The improved performance that \herschel brings to CEP analyses  is assessed in Sect.~\ref{sec:physics}.  The role of \herschel in the LHCb trigger is explained in Sect.~\ref{sec:hrcInTrigger}. Conclusions are drawn in Sect.~\ref{sec:conclusions}.

\section{Physics motivation: CEP studies at LHCb}
\label{sec:motivation}

LHCb is a forward spectrometer with tracking, particle identification and calorimetry extending to  a pseudorapidity of $\eta \sim 5$~\cite{LHCb-DP-2014-002}. The upstream stations  of the silicon-strip vertex detector (VELO) provide sensitivity to charged particles down to  $\eta \sim -3.5$  in the backward hemisphere. 
Although the design of LHCb was optimised for studies of heavy-flavour hadron decays, its operating conditions and sensitivity to low transverse-momentum (\pt) particles  make it ideally suited to measurements of CEP processes.

\begin{figure}[htp]
\begin{center}
\includegraphics[width=.99\textwidth]{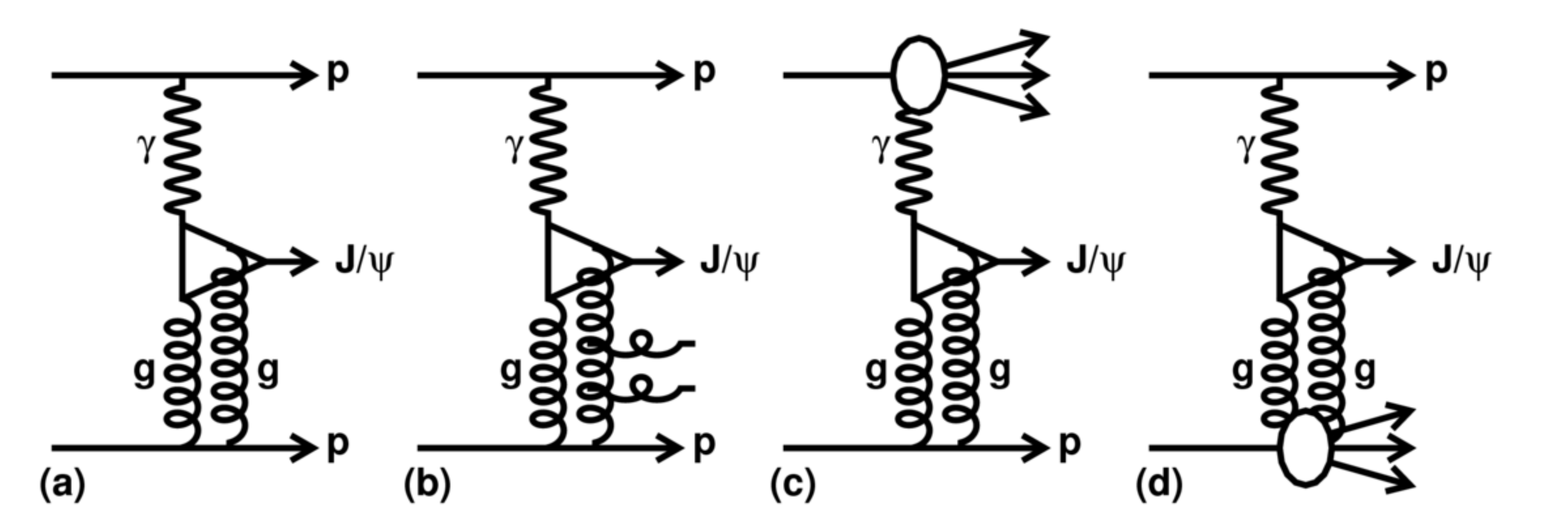}
\caption{Feynman diagrams of diffractive-production mechanisms of $J/\psi$ mesons at the LHC, where the double gluon system being emitted from the beam proton constitutes the pomeron.  (a) is the pure CEP process,  (b) has additional gluon radiation, and (c) and (d) involve proton dissociation.   (Taken from Ref.~\cite{LHCb-PAPER-2012-044}).}
\label{fig:feynman}
\end{center}
\end{figure}

In CEP interactions at the LHC a central system of one or more particles is formed as the result of pomeron or photon exchange. The beam protons survive the collision intact,  remaining undetected inside the beampipe.  As an example, the Feynman diagram for the CEP formation of a $J/\psi$ meson is shown in Fig.~\ref{fig:feynman}.
The low multiplicity of the final state and the absence of activity, or large `rapidity gap', either side of the central system provides a distinctive signature of the CEP process. 
In practice, this signature can only be exploited by the current LHCb trigger in beam crossings where there are no additional interactions.  The fraction of single interaction events within the acceptance of the LHCb spectrometer is relatively large during normal operation, being approximately 37\% in 2015.
 Contamination to the signal selection arises, however, from inelastic interactions in which one or both protons disassociate, or gluon radiation accompanies the pomeron or photon exchange, processes that are also displayed in Fig.~\ref{fig:feynman}.   In these cases additional  hadrons are produced at high rapidities, which generally means that they fall outside the acceptance of the sub-detectors  of the spectrometer.  

LHCb has published studies of exclusive $J/\psi$, $\psi(2S)$~\cite{LHCb-PAPER-2014-027,LHCb-PAPER-2013-059,LHCb-PAPER-2012-044} and $\Upsilon$~\cite{LHCb-PAPER-2015-011} production based on the Run-1 data set.  The background from inelastic events in which high rapidity particles were undetected in the spectrometer was estimated by fits to the $\ptsq$ spectra of the candidates.  The expected distribution of CEP signal peaks at $\ptsq \sim 0$, although the exact distributions of signal and non-exclusive background depend on model-based assumptions.  Using this approach, the typical purity of the selection was determined to be $50-60\%$.  

The sensitivity that \herschel provides for high-rapidity particles will enable the contamination from inelastic events in the CEP selection to be suppressed. Information from \herschel will typically be deployed in a veto mode.  The absence of any significant activity in the FSCs will be used to confirm the existence of a rapidity gap extending beyond the spectrometer acceptance, and to add confidence to the central-exclusive hypothesis of CEP candidates.   In this manner the purity of the CEP selection will  be improved, and the systematic uncertainties associated with the modelling of the residual background can  be reduced.

The information provided by \herschel will be beneficial for other analyses in which rapidity gaps are an important signature, for example the study of single-diffractive events. It can also be used as a counter of general inelastic activity, which has applications in the luminosity determination.

\section{Description of the detector}
\label{sec:setup}

\subsection{Layout}
\begin{figure}
  \includegraphics[width=.95\textwidth]{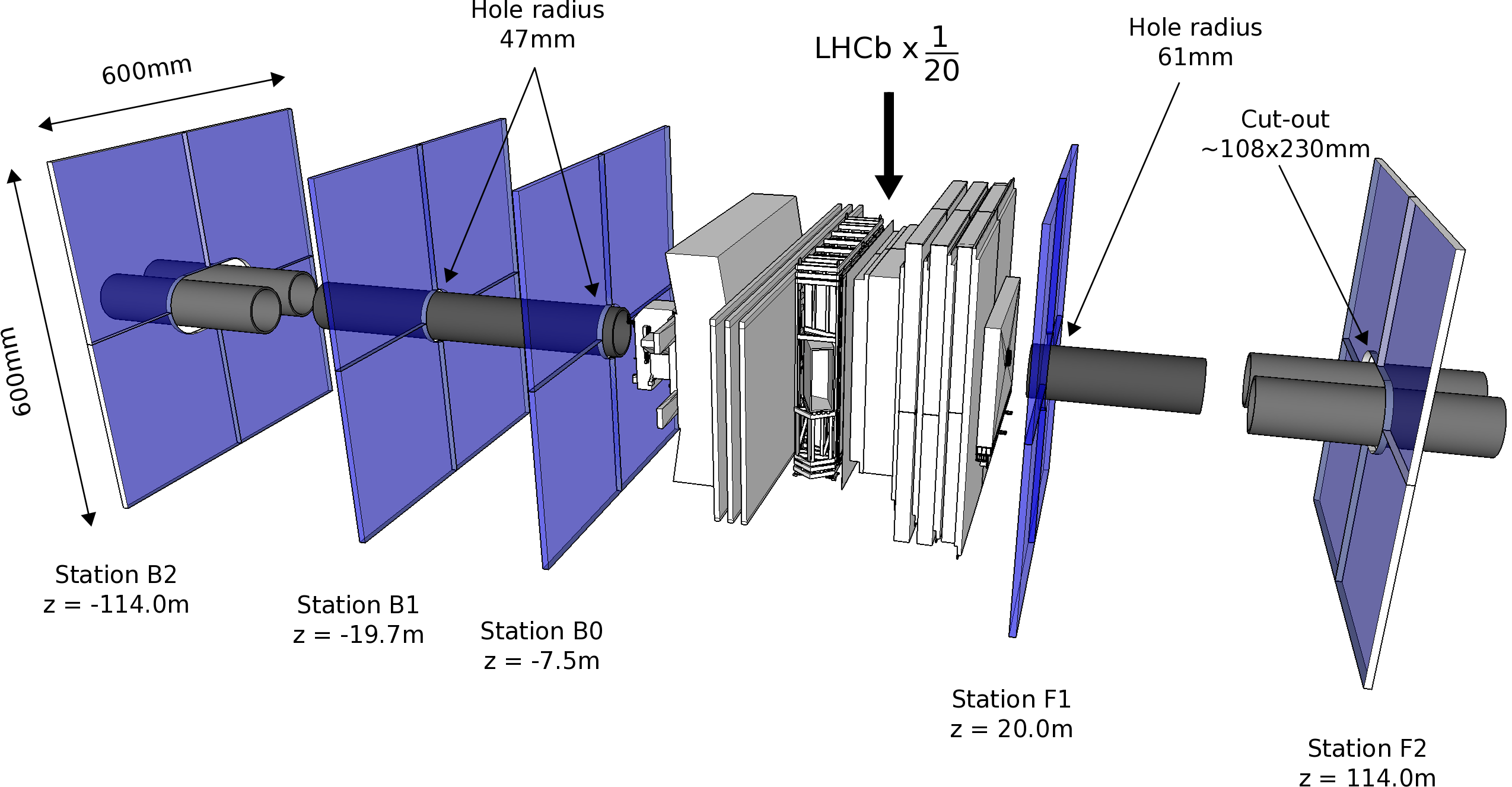}
  \caption{Layout of the active areas of the \herschel stations around the LHCb interaction point (IP8), where for illustration the \herschel stations have been magnified by a factor of 20 with respect to the rest of the LHCb detector. $z$-axis not to scale.}
  \label{Fig:LayoutIR8}
\end{figure}
In the LHCb coordinate system
the $z$-axis points from the LHCb interaction point (IP8) 
towards the muon chambers and is collinear with the nominal beam line.
As shown schematically in Fig.~\ref{Fig:LayoutIR8}, the \herschel system comprises three stations at negative $z$, known as `backward' or `B' stations, and two stations at positive $z$, known as `forward' or `F' stations.

The station closest to the interaction point, named `B0', is located at $z\sim-7.5$\,m, next to the MBXWH compensator dipole. The next closest stations are `B1' and `F1', located at $z\sim-19.5$\,m and $z\sim20$\,m, respectively, in the proximity of the MBXWS corrector dipoles. The most distant stations, `B2' and `F2', are located at $z\sim\pm114$\,m, close to the point at which the vacuum chamber splits into two separate chambers, one for each beam.

The active element of each station is a plastic scintillator plane with 
outer dimensions of $600$\,mm\,$\times$\,600\,mm, centred around the beam line.
The shape and dimensions of the inner opening depend on
the local vacuum chamber layout.
Stations B0, B1, and F1 have circular holes with radii of 47\,mm (B0, B1) and 
61\,mm (F1), respectively.
For stations B2 and F2, the inner opening has a half-width of 115\,mm 
in the horizontal direction (to encompass the two vacuum chambers), and 
a half-width of 54\,mm in the vertical direction.
\subsection{Acceptance}
Stations B0, B1, and F1 are intended to register the showers 
produced by high-rapidity collision products crossing the 
beam pipe inside or close to the MBXWH and MBXWS magnets 
(which constitute local aperture restrictions).
Stations B2 and F2 are intended to detect showers from high-rapidity 
neutral particles interacting with the copper absorber block of the 
LHC collision rate monitors (BRAN~\cite{Matis:2016raz}),
which are located close to and in front of these \herschel stations.

In order to validate the detector concept and to determine the acceptance,
the expected activity in the \herschel stations was simulated 
using the \gauss simulation framework \cite{LHCb-PROC-2011-006}, 
generating minimum bias events using \pythia~8 \cite{Sjostrand:2007gs} 
and transporting the emerging final-state particles through the detector 
and tunnel geometry using \geant \cite{Agostinelli:2002hh,Allison:2006ve}.

\begin{figure}
  \centering
  \includegraphics[width=0.95\textwidth]{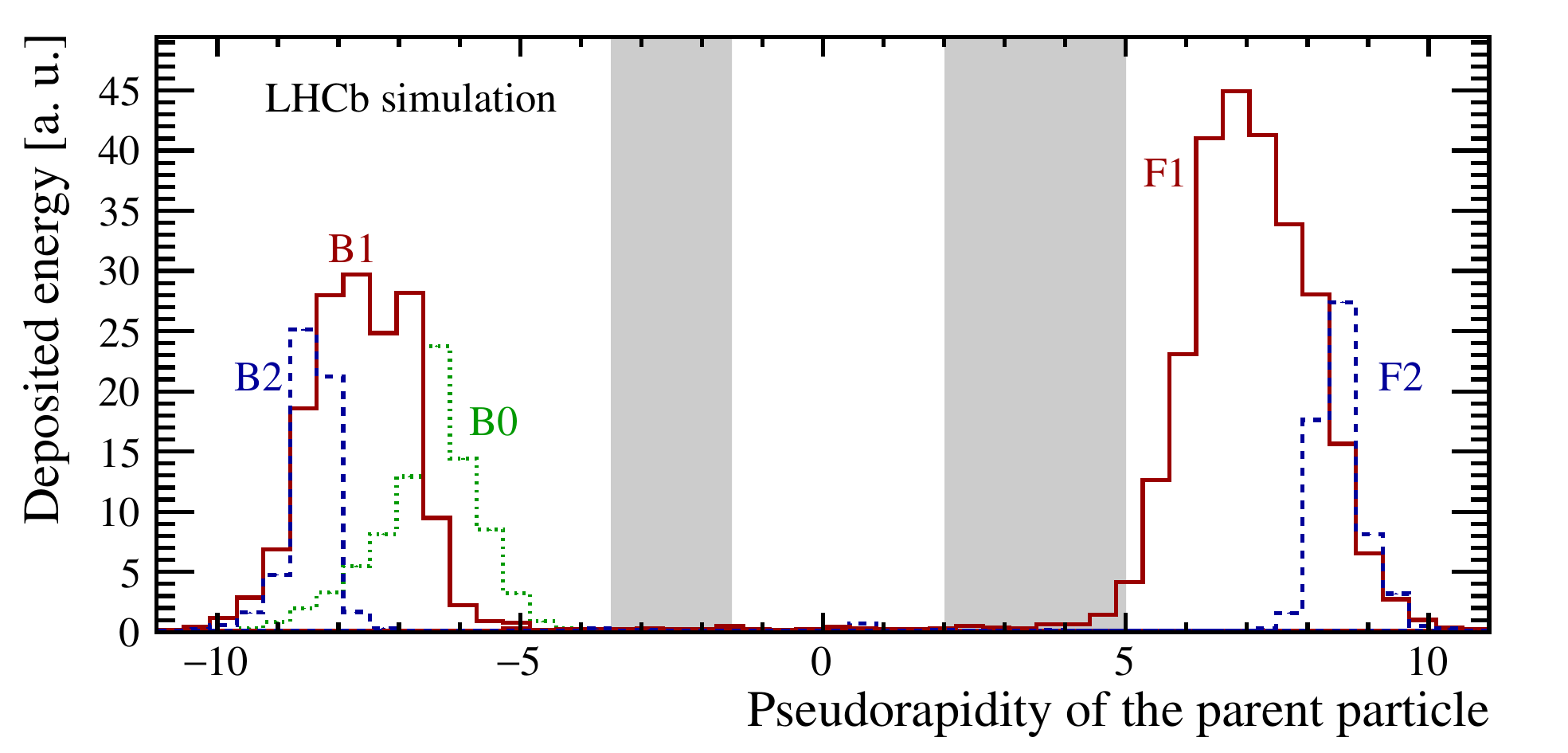}
  \caption{Energy deposit in the scintillators as a function of the 
           pseudorapidity of the parent particle 
           that caused the shower. The grey areas indicate 
           the nominal pseudorapidity coverage of LHCb.}
  \label{Fig:SimEtaMinBias} 
\end{figure}

The simulations confirm that the signal in the \herschel stations is dominated by 
energy deposits from low-energy electrons and positrons produced in 
the showers induced by high-rapidity particles interacting with machine 
elements close to the scintillators. 
Fig.~\ref{Fig:SimEtaMinBias} illustrates the angular coverage, showing the pseudorapidity of the particles produced in $pp$ collisions where the contribution from each particle is weighted by the corresponding energy deposit in the scintillators.  The asymmetry between the distributions for B1 and F1 
           is a consequence of the asymmetric beampipe layout on the two sides of the LHCb interaction point, leading also to the different hole radii for the B1 and F1 counters shown in Fig.~\ref{Fig:LayoutIR8}. The \herschel stations are seen to add detector 
acceptance from five up to nearly ten pseudorapidity units.

\subsection{Scintillators and photodetectors}
\label{sec:detector}
\begin{figure}[!htb]
  \centering
  \includegraphics[height=0.3\textheight]{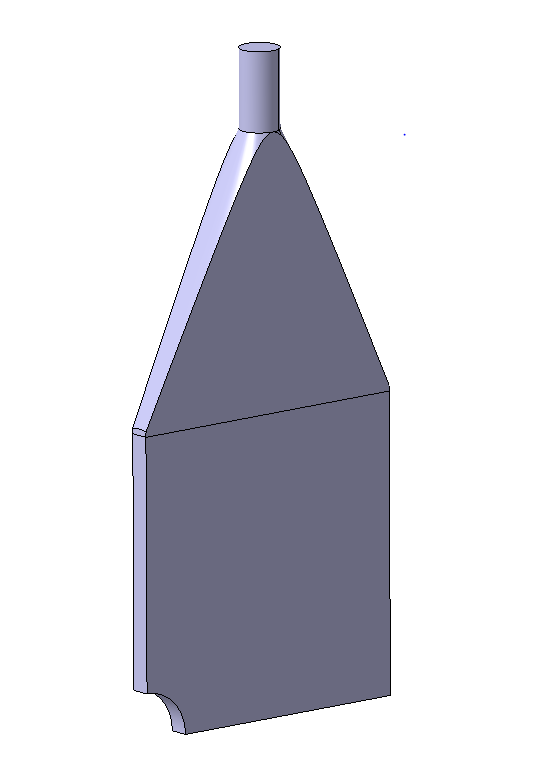}
  \hspace*{5em}
  \includegraphics[height=0.3\textheight]{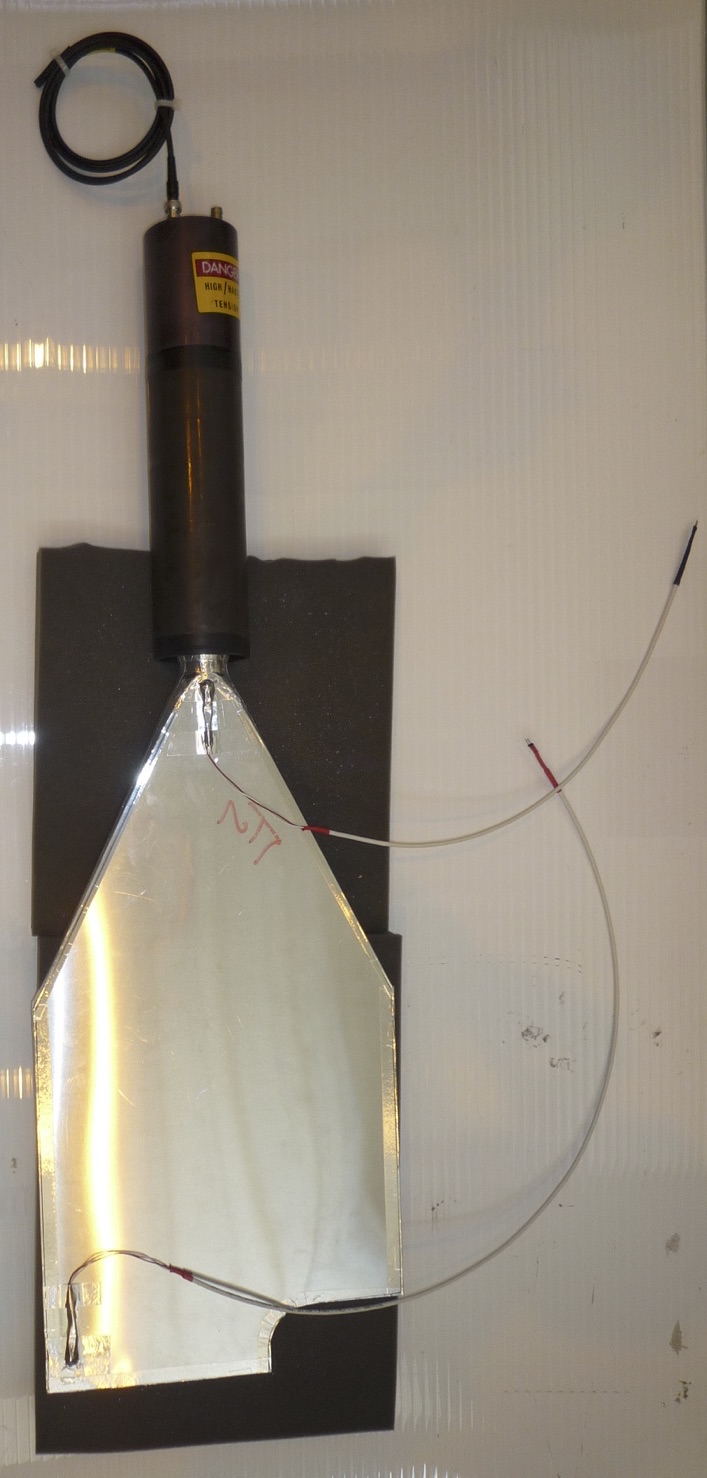}
\caption{Schematic design of the scintillator and light guide of a single quadrant 
of B0/B1-type (47\,mm inner radius), accompanied by a photograph of a B2/F2-type quadrant with the PMT and 
`clipping cable' attached.}
\label{fig:counter}
\end{figure}

Each station is segmented into four quadrants, \ie four 
scintillator plates read out by separate PMTs.
The scintillator plates are 20\,mm thick and are manufactured from
EJ-200\footnote{Eljen Technology, Sweetwater, Texas, 
United States (\url{http://www.eljentechnology.com}).} 
plastic scintillator material, with a rise time of 0.9\,ns and 
a decay time of 2.1\,ns. 
Plexiglass light guides of fishtail type are glued to one side of 
the scintillator plates, providing a transition between the 
rectangular cross-section of the scintillators 
and the circular cross-section of the PMT (Fig.~\ref{fig:counter}, left). 
For safety reasons\footnote{Polyvinyltoluene emits dense black smoke when 
burning without sufficient oxygen.} and to prevent external light leaking in, 
the scintillators and light guides are covered by thin aluminium sheets.

Each light guide is coupled to a Hamamatsu\footnote{Hamamatsu Photonics K. K., Hamamatsu City, Japan (\url{www.hamamatsu.com}).} R1828-01 $2^{''}$ PMT.
This PMT is a suitable choice because of its 
relatively fast rise time of 1.3\,ns and its large range of gain adjustment.

A general requirement for the stable operation of a PMT is that the 
resistive divider current exceed the average anode current by a factor of 
at least 10 -- 20. 
The R1828-01 PMT is rated for a maximum average anode current of $200\,{\rm \upmu A}$, and the voltage divider should thus drain about $2-4$~mA at the relatively low gain 
($\sim 10^3-10^4$) used in operation. 
This is achieved by a custom-made divider that provides 
an extra bias current to the last set of dynodes, 
allowing the initial current through the divider to be restricted, 
while ensuring that there is enough current provided in the vicinity 
of the anode during nominal operation. 

The use of Zener diodes for some of the stages allows a stable voltage to be maintained for the last dynodes at a high current drain. 
The two voltages for each PMT (bias and high voltage) are supplied from
CAEN\footnote{CAEN S. p. A., Viareggio, Italy (\url{http://www.caen.it/}).}  
A1535 and A1538D modules located on the accessible side of the LHCb cavern. 

A short coaxial cable terminated with a resistor-capacitor chain 
is connected in parallel to the signal cable at the output of the PMT 
to `clip' the tail of the signal and thus 
ensure that the pulse width is less than 25\,ns.
A complete assembly of scintillator, light guide and PMT 
is shown in Fig.~\ref{fig:counter} (right).

The PMT signals are transmitted to the readout electronics located in 
the LHCb cavern using low-loss coaxial cables (C-50-11-1).
The length of these cables ranges between 137\,m (for station B2), 
corresponding to a signal attenuation of 15\%, and 15.4\,m (for station B0), 
corresponding to an attenuation of 5\%. 
\subsection{Mechanics}
\begin{figure}[!htb]
  \centering
  \includegraphics[width=0.47\textwidth]{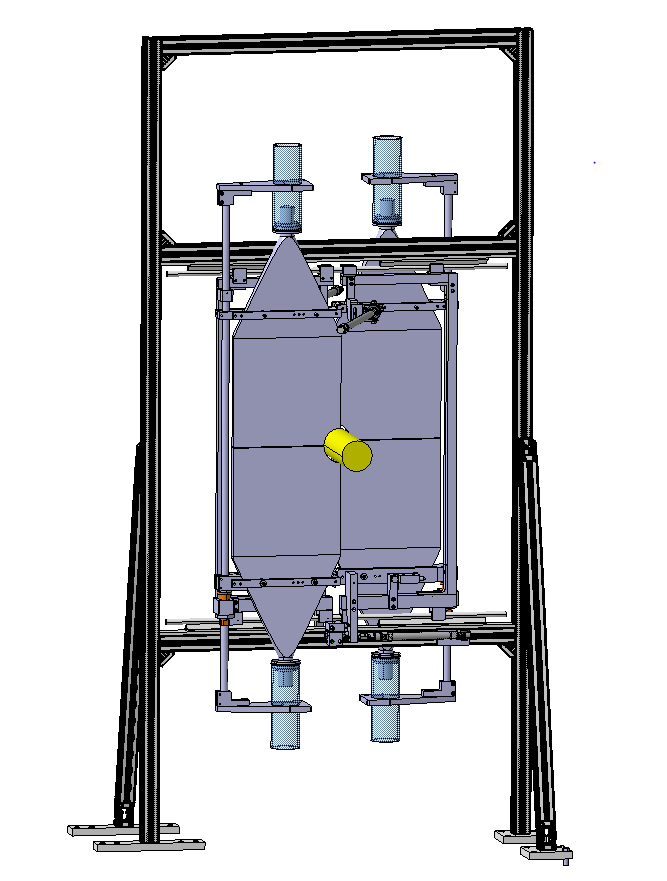}
  \includegraphics[width=0.47\textwidth]{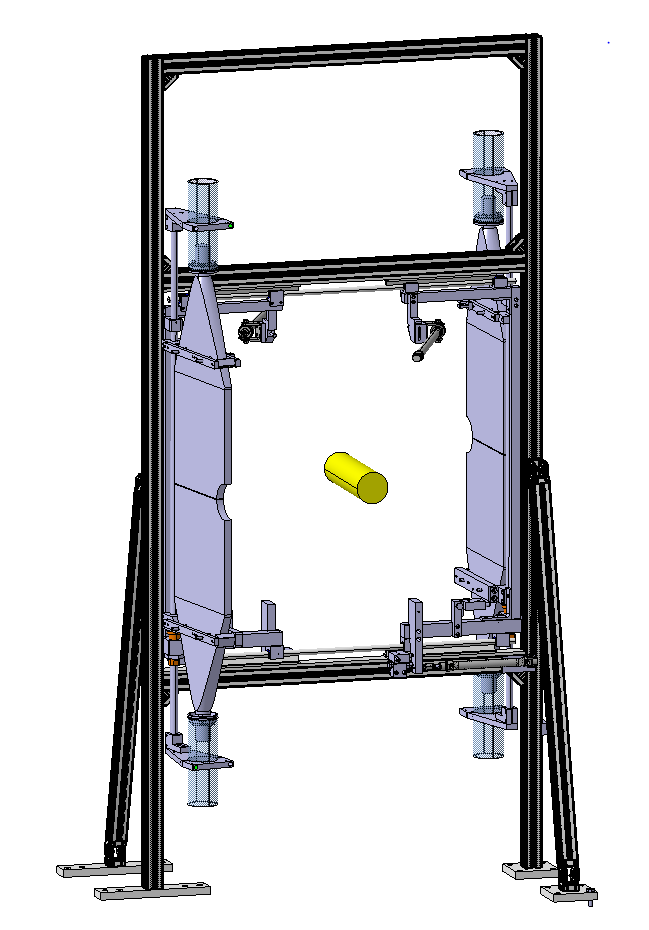}
  \caption{Visualisation of a \herschel station in its nominal data-taking position and in its parking position, when the detectors are retracted and rotated.}
\label{fig:openandclosed}
\end{figure}

\begin{figure}[htb]
  \centering
  \includegraphics[angle=-90,width=0.32\textwidth]{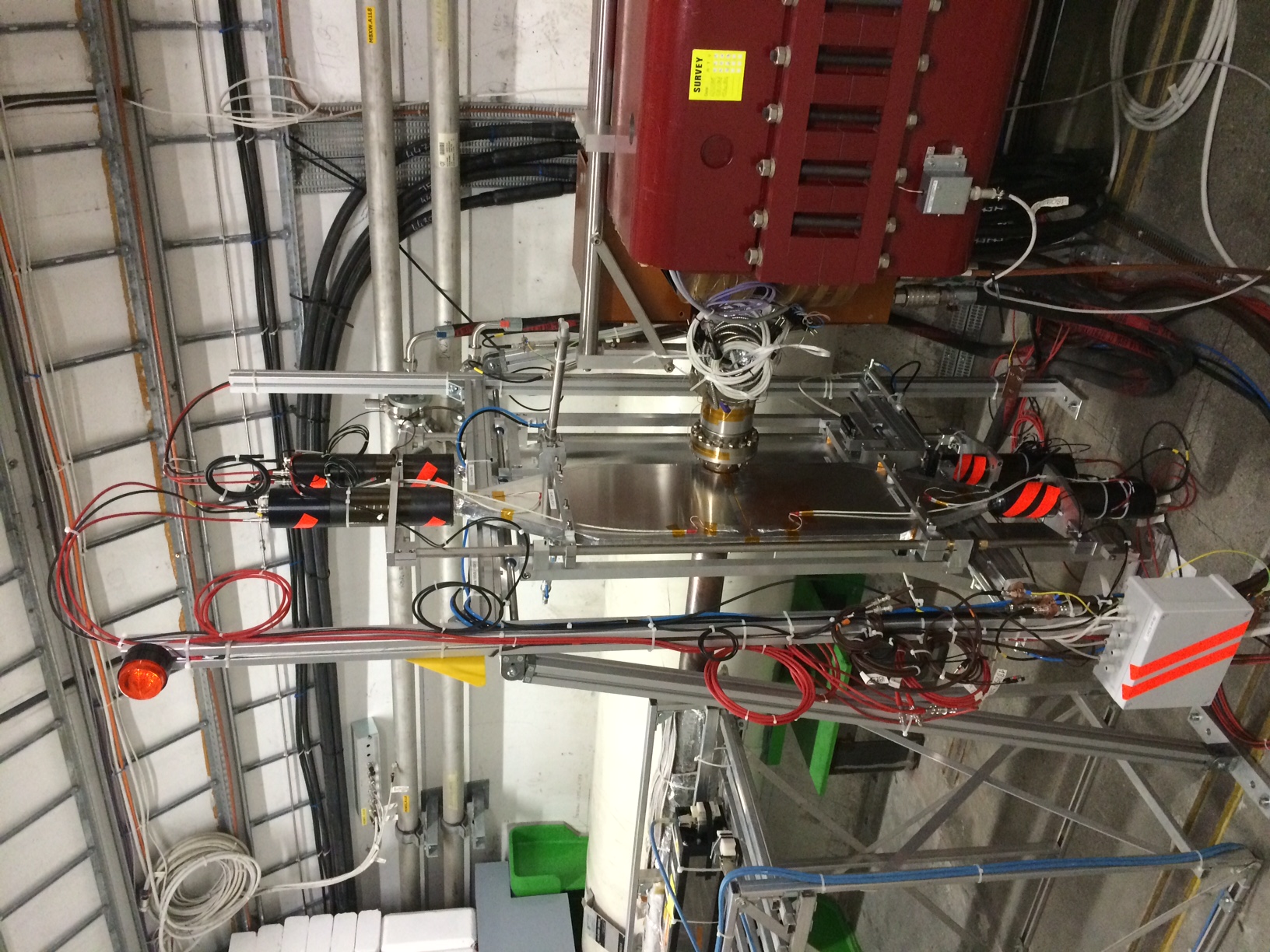}
  \includegraphics[angle=-90,width=0.32\textwidth]{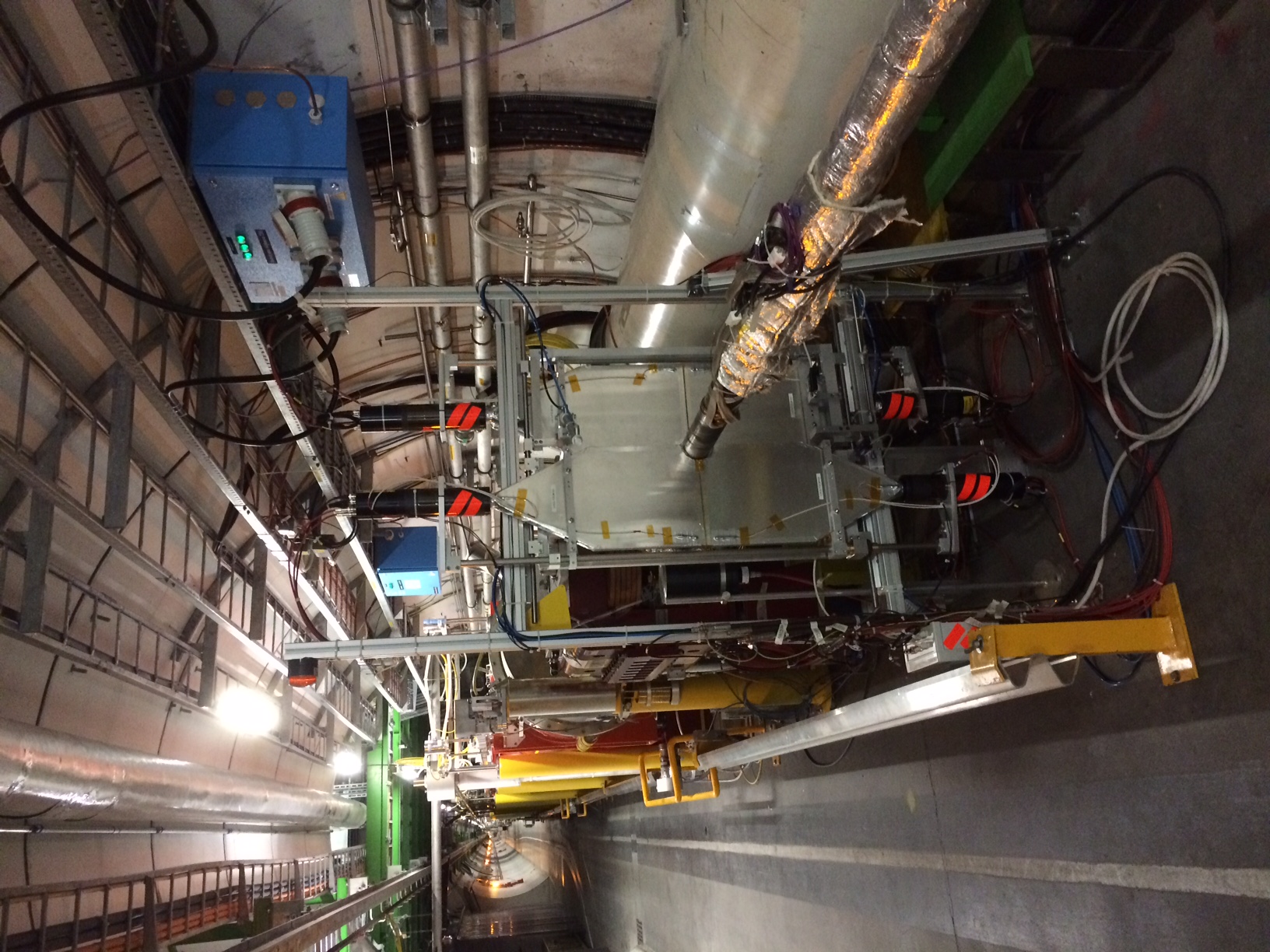}
  \includegraphics[angle=-90,width=0.32\textwidth]{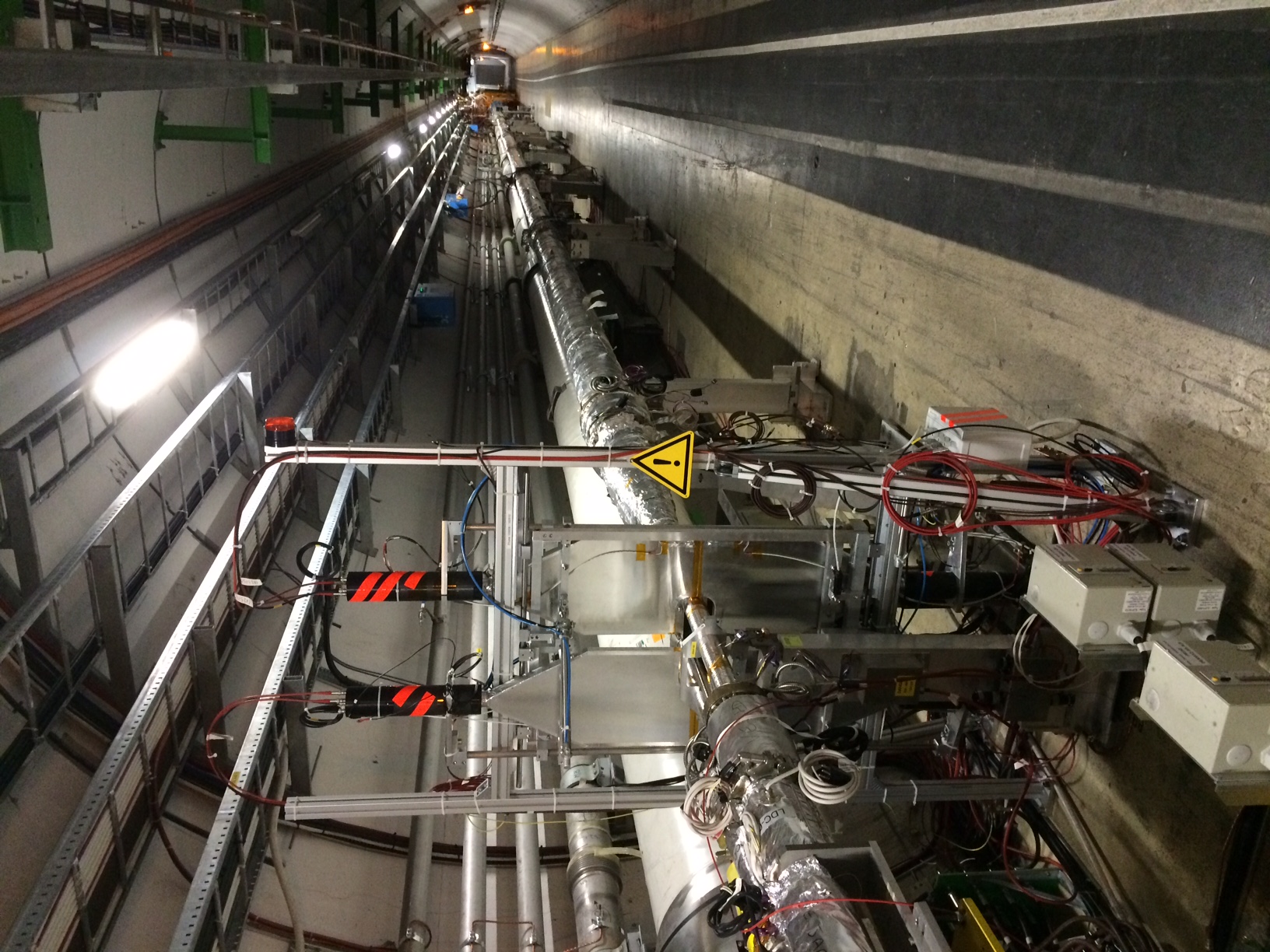}
  \caption{Photographs of the backward \herschel stations B0, B1, and B2, respectively.}
  \label{fig:integration_b}
\end{figure}

\begin{figure}[htb]
  \centering
  \includegraphics[angle=-90,width=0.3\textwidth]{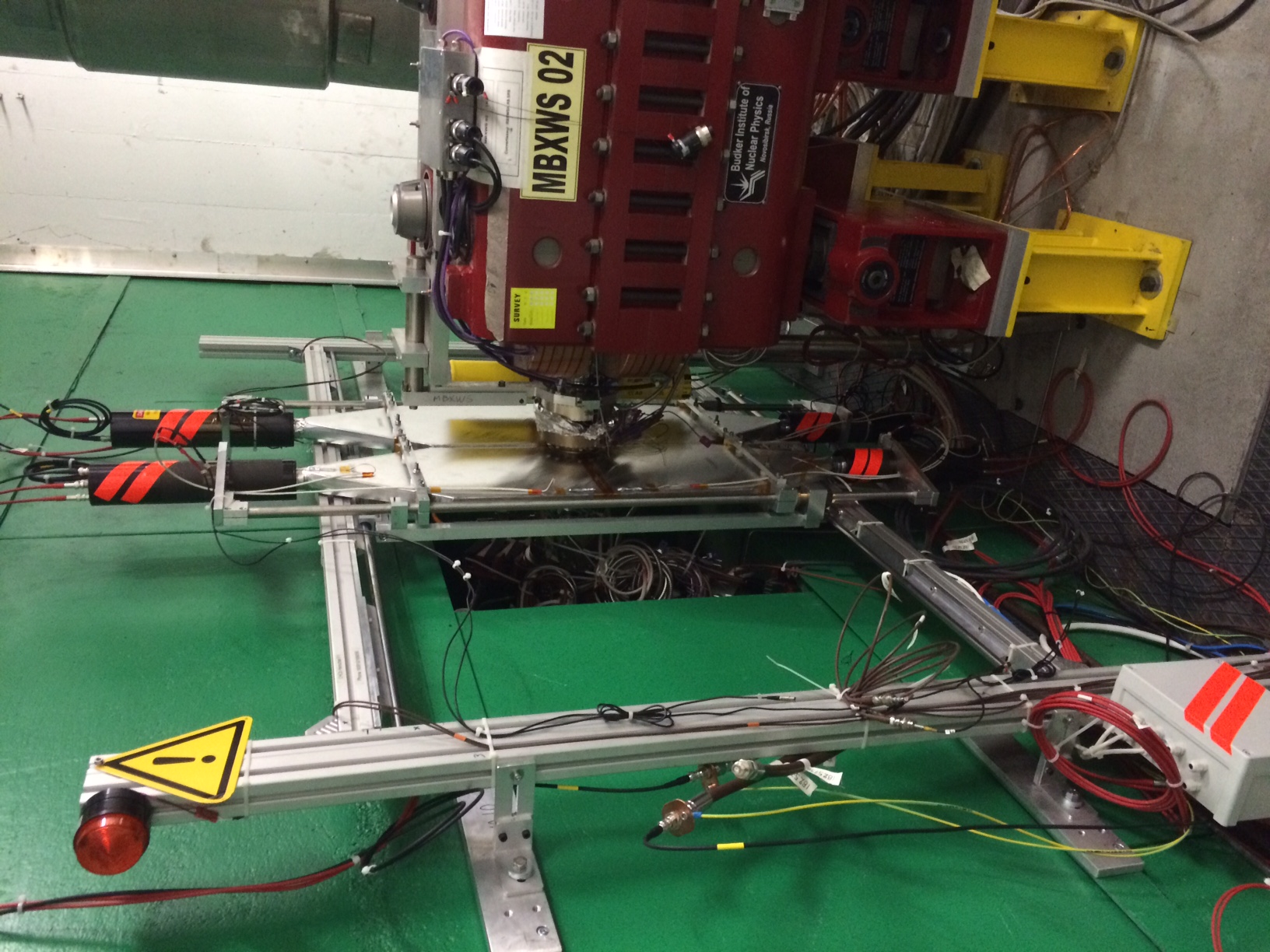}
  \includegraphics[angle=-90,width=0.3\textwidth]{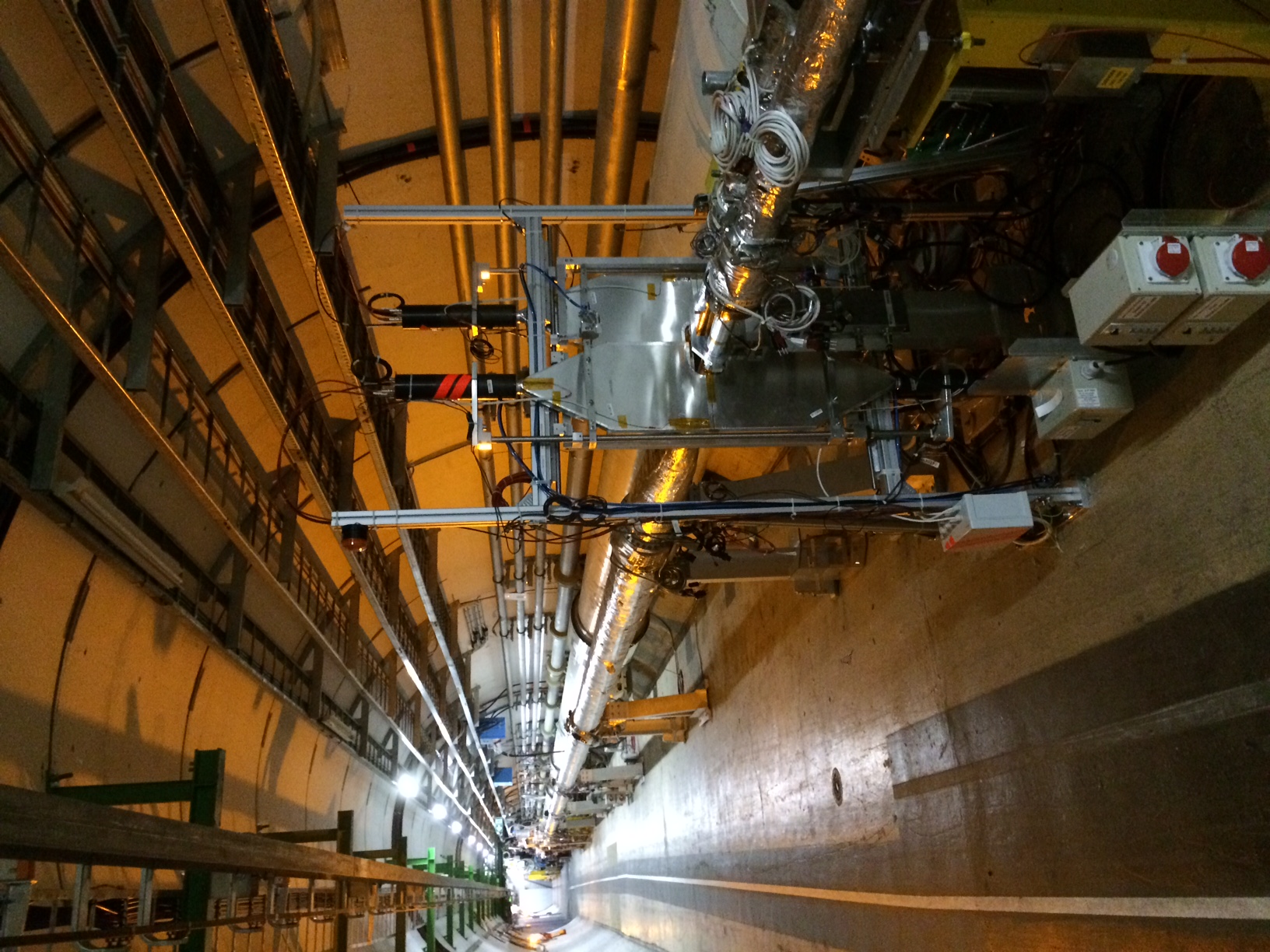}
  \caption{Photographs of the forward \herschel stations F1 and F2, respectively.}
  \label{fig:integration_f}
\end{figure}

Each photodetector is mounted in a steel housing, 
including a shielding tube that provides protection 
against magnetic fields up to approximately 0.1\,T.
The detectors are mounted on mechanical supports manufactured from standard aluminium Bosch profiles. 
The stations are equipped with a remote-controlled, pneumatic motion system, illustrated in Fig.~\ref{fig:openandclosed}, 
which allows the scintillators to be retracted from the high-fluence region 
close to the vacuum chamber whenever they are not required for
data-taking for an extended period of time.
Figures~\ref{fig:integration_b} and \ref{fig:integration_f} show photographs of the 
shower counter stations as installed in the LHC tunnel. 

\subsection{Readout and data acquisition}
\label{sec:detector:readout}
\begin{figure}
  \centering
  \includegraphics[width=.7\textwidth]{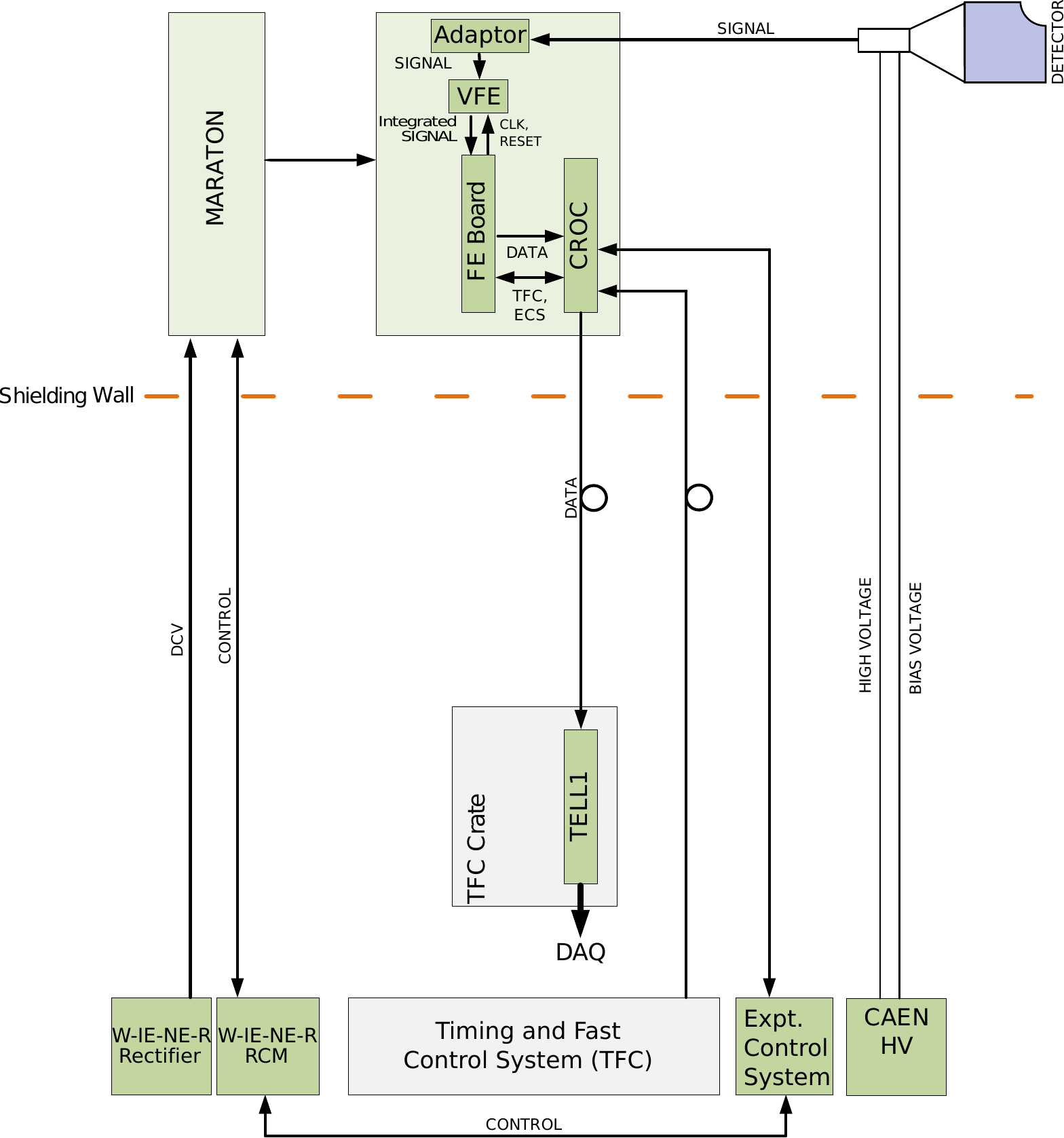}
  \caption{Block diagram of the readout, powering, and control of the FSC detectors. The diagram shows only one quadrant and one of the two front-end crates.}
  \label{fig:readoutoverview} 
\end{figure} 
In order to minimise the development cost and to ease the integration 
with the LHCb data acquisition and trigger system, 
\herschel makes use of the electronics of the 
LHCb preshower system \cite{LHCb-2003-038}, comprised of
a very-front end (VFE) board, a front-end board (FEB), 
and a calorimeter readout card (CROC).
The FEB and CROC are housed in a front-end crate 
(VME 9U card cage with a
W-IE-NE-R\footnote{W-IE-NE-R, Plein \& Baus, Burscheid, Germany (\url{www.wiener-d.com})} 
 MARATON power supply and a backplane specific to the LHCb calorimeter system). 
Two front-end crates are installed in the LHCb cavern, one for each side.  Fig.~\ref{fig:readoutoverview} gives an overview of the readout chain 
components and connections.

The multi-anode photomultipliers of the preshower system are connected
directly to the VFE board, whereas in \herschel the signals are transmitted over long coaxial 
cables. A dedicated adapter board was therefore developed for \herschel 
which receives the signal from the PMTs and adapts it to the input requirements 
of the VFE board.
Each channel of the VFE board has two multiplexed differential integrators running at 20\,MHz allowing the VFE channel reading even-numbered bunch crossings an empty, relaxation period while the other channel reads out, and vice-versa.
The integrated analogue signal is sent to the FEB
which digitises it with a 10-bit ADC and distributes it to the data acquisition and trigger paths.
The digitised data are stored in looping RAMs with a depth of 256 cycles, 
in groups of four channels. 
The fine delay of the integration and sampling clock can be adjusted 
in steps of 0.78\,ns in order to maximise the signal and minimise the contamination from signal in $pp$ crossings preceding and following the one of interest. 

On a positive decision from the global LHCb hardware trigger (L0),
the data from each station are sent from the corresponding RAM depth. 
This allows for the data from the different stations, 
which acquire up to 750\,ns relative delays due to the different station positions, to be synchronised with the global LHCb bunch counter.
The digitised signals are sent to the CROC via the crate backplane.
The CROC plays a dual role. Firstly, it is responsible for the control and monitoring
of the front-end crate. 
It receives the timing and fast control (TFC) information 
and distributes it on the backplane to the FEB. 
The interface to the Experiment Control System (ECS) is based on the SPECS 
protocol \cite{LHCb-2003-004}.  
The ECS control and monitoring of the FEBs  
is communicated through the backplane. 
Secondly, the CROC gathers the front-end data and sends them to the 
acquisition board (TELL1 \cite{Haefeli2006494}) via optical fibres.

\section{Detector sensitivity}
\label{sec:characterisation}
In the case of a single CEP interaction, the produced central system may be reconstructed inside the standard LHCb acceptance and will be accompanied by no other activity within the main detector. For this signal no activity is expected in the \herschel system. In the case of non-exclusive processes additional particles are produced together with the CEP candidate. These particles lead to activity being registered in \herschel. In this Section the sensitivity of \herschel to charged-particle activity is considered. Steps taken to minimise noise in order to maximise sensitivity are described. Finally the characteristic response of the detector in the absence of additional particle activity is presented. This `empty detector' signal is the expected response in the case of a single CEP interaction in a $pp$ bunch crossing.

\subsection{Response to incident charged particles}
\begin{figure}[!htb]
  \centering
  \includegraphics[width=0.48\textwidth]{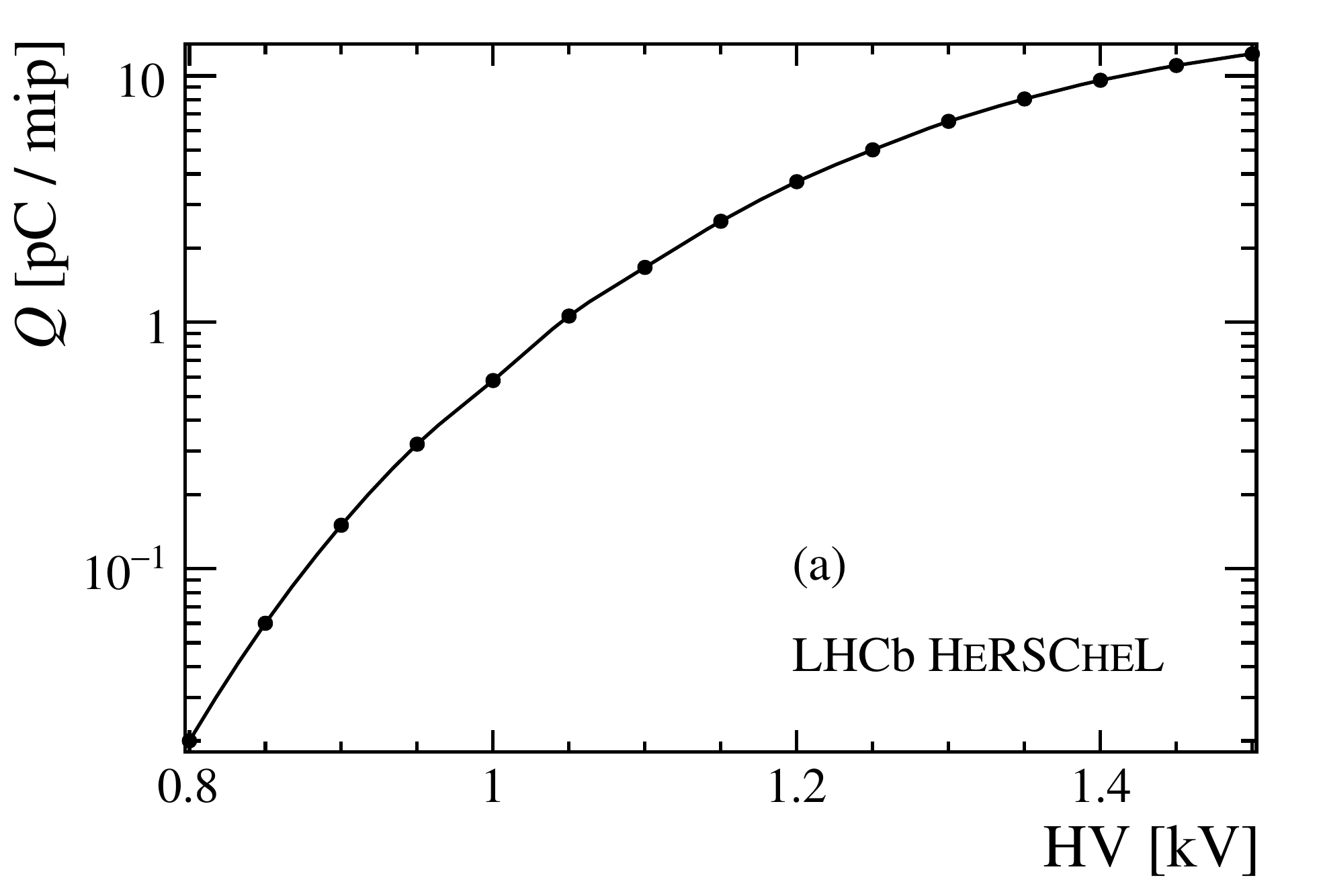}
  \includegraphics[width=0.48\textwidth]{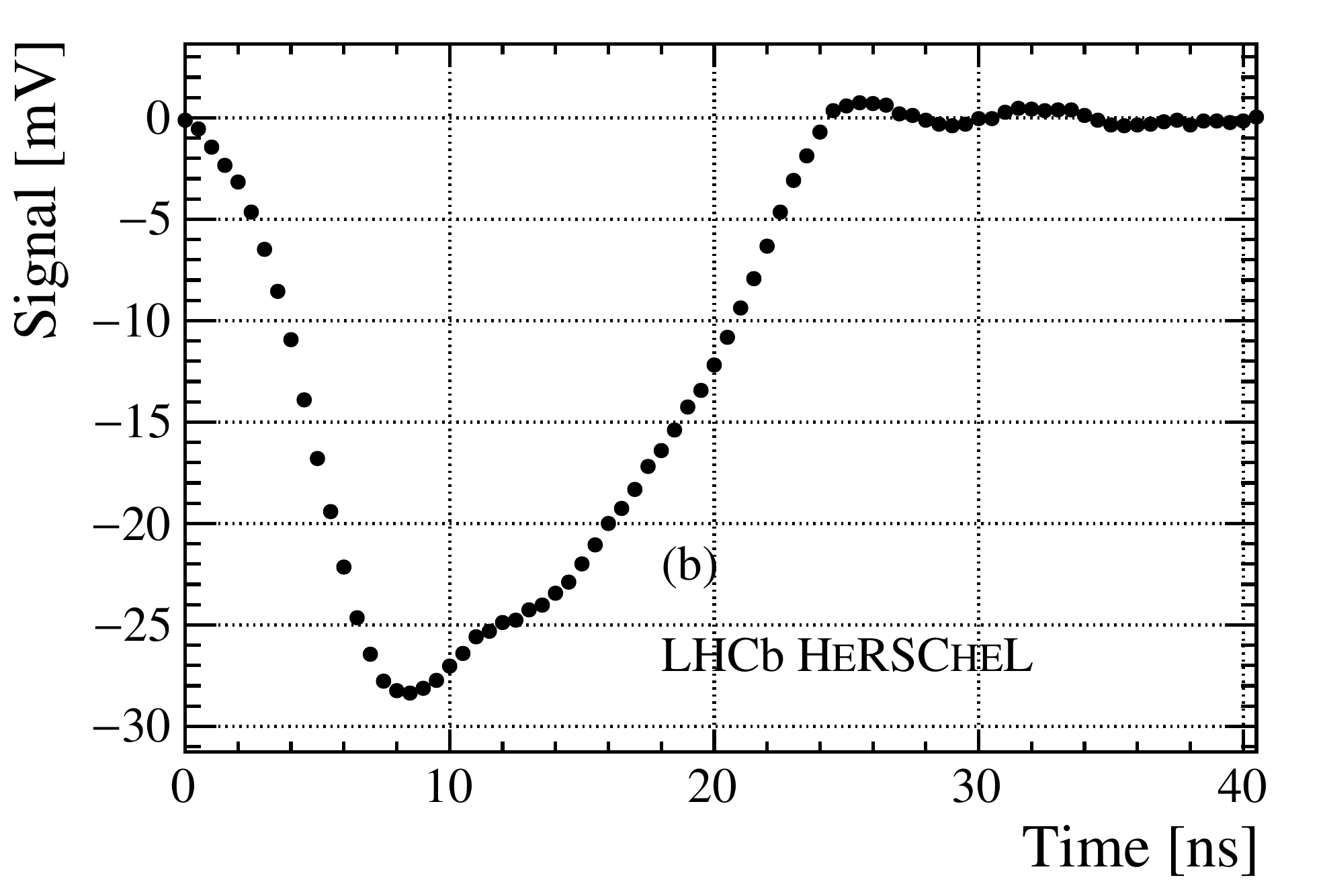}
\caption{
(a) shows the signal per particle as a function of PMT high voltage for one of the quadrants of station B0. The absolute scale is determined using measurements with cosmic muons, and the evolution with high voltage using a pulsed LED setup. The analogue signal of one quadrant of station B0, recorded during injection tests (`TED shots') in November 2014, is shown in (b).} 

\label{fig:calib}
\end{figure}

The detector response to charged particles has been calibrated in the laboratory, prior to installation, using cosmic muons. To achieve sensitivity to single incident particles,
the voltage applied to the PMT was set to 1.1\,\kV. Later, the PMT gain as a function of the applied high voltage has been measured using LED light pulses. Using these measurements 
the signal per charged particle is determined as a function of the applied high voltage. This is shown in Fig.~\ref{fig:calib}(a) for one quadrant of station B0. 

Figure~\ref{fig:calib}(b) shows the response (analogue pulse shape) of one 
of the quadrants of station B0, 
recorded using a digital oscilloscope during LHC injection tests in November 2014.
In these tests, proton beams were extracted from the Super Proton Synchrotron (SPS) and dumped on an 
absorber (named TED) located approximately 340\,m away from the LHCb 
interaction point, generating a shower of secondary particles at small angles 
with respect to the beam line. 

At the beginning of the 2015 data-taking period, the high-voltage settings were 
configured such that 
an energy deposit in the scintillator corresponding to one minimum ionising 
particle (MIP) resulted in a digitised signal between two and five ADC counts. 
These settings were kept throughout 2015.

\begin{figure}[htb]
  \centering
  \includegraphics[width=0.55\textwidth]{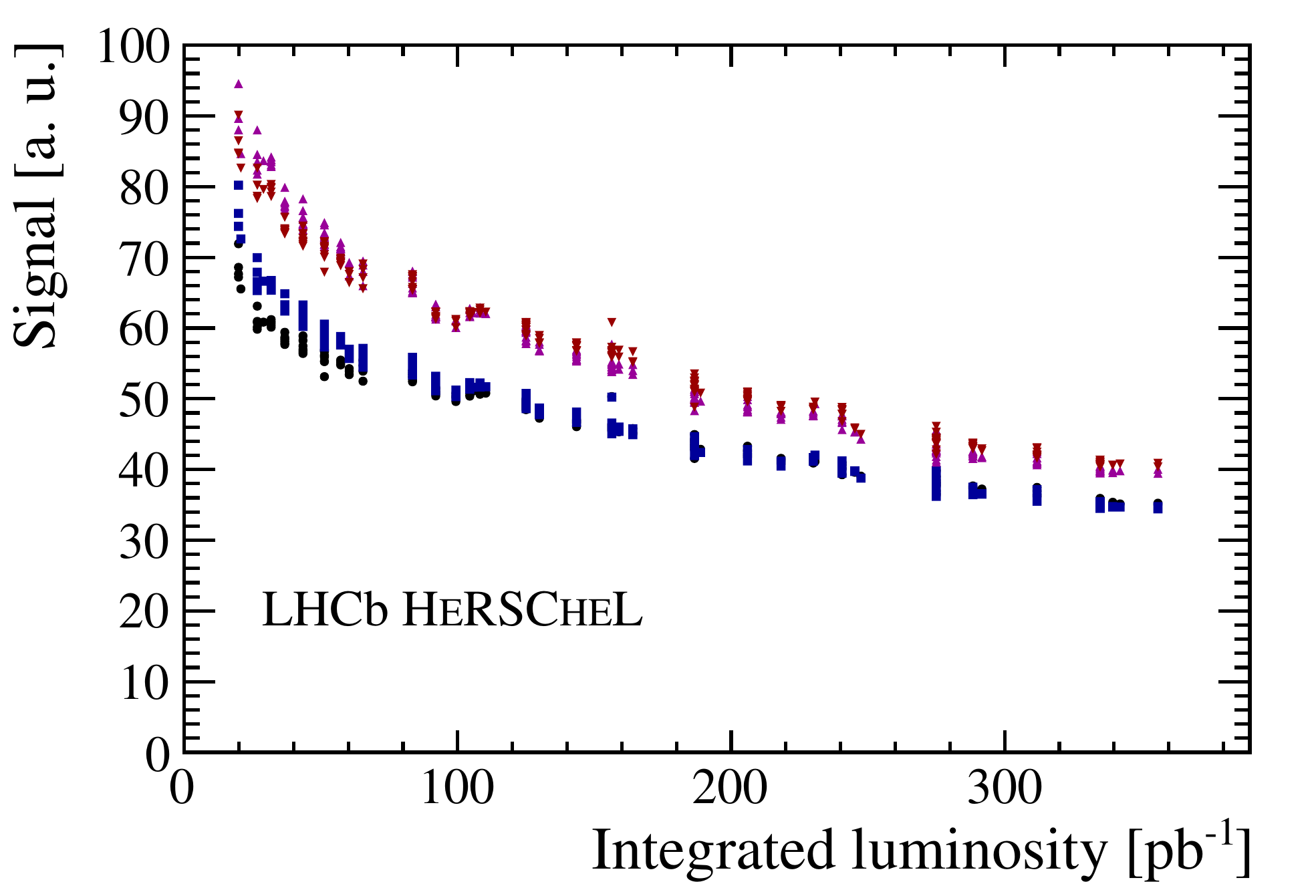}
  \caption{Average signal in $pp$ collisions
           as a function of integrated luminosity in 2015, 
           for the four quadrants of station B0. The high-voltage settings 
           were kept the same throughout the entire 2015 data-taking period.}
  \label{fig:ageing}
\end{figure}

Irradiation leads to a reduction in attenuation length of scintillators and light guides and consequently 
to a loss of signal. In order to quantify the average signal, the tail of the ADC spectrum above the pedestal (see, for example, Fig.~\ref{im:Pedestals})
is fitted with an exponential distribution, and the mean of the fitted  
function is used as a metric of the average signal.
The ageing process is illustrated in Fig.~\ref{fig:ageing}
which shows the average signal
per $pp$ collision in the quadrants of station B0 as a function 
of the integrated luminosity in LHCb. The PMTs of this station 
were operated at a high voltage of $830 - 850$\,V, depending on the quadrant. 
The average signal corresponded to 
the equivalent of $\sim72 - 95$ MIPs at the
beginning of the data-taking period and degraded to the equivalent of 
$\sim35 - 40$ MIPs at the end of 2015.

In 2016 and 2017, the high-voltage settings were updated regularly to compensate for the 
increasing loss of photoelectrons due to radiation damage.  
For example, the high voltages of station B0 were set to $\sim830 - 860$\,V at the beginning of the 
2016 data-taking period and were raised gradually up to $\sim880 - 920$\,V at the end of 2016. 
The increase in high voltage did not lead to a significant change of the dark count rate. 


\subsection{Noise}
\label{sec:PedCalib}
For data collected during 2015 a significant common-noise component was found between the detector channels within each station, arising either from interference induced in the cables passing along the tunnel, or in the front-end electronics. This common noise has been studied by considering the correlation between the signal from each input channel and one spare cable for each station, placed next to the signal cables in the tunnel and connected to the readout electronics but left unconnected at the detector end.

\begin{figure}[htbp]
\begin{center}
\includegraphics[width=.49\textwidth]{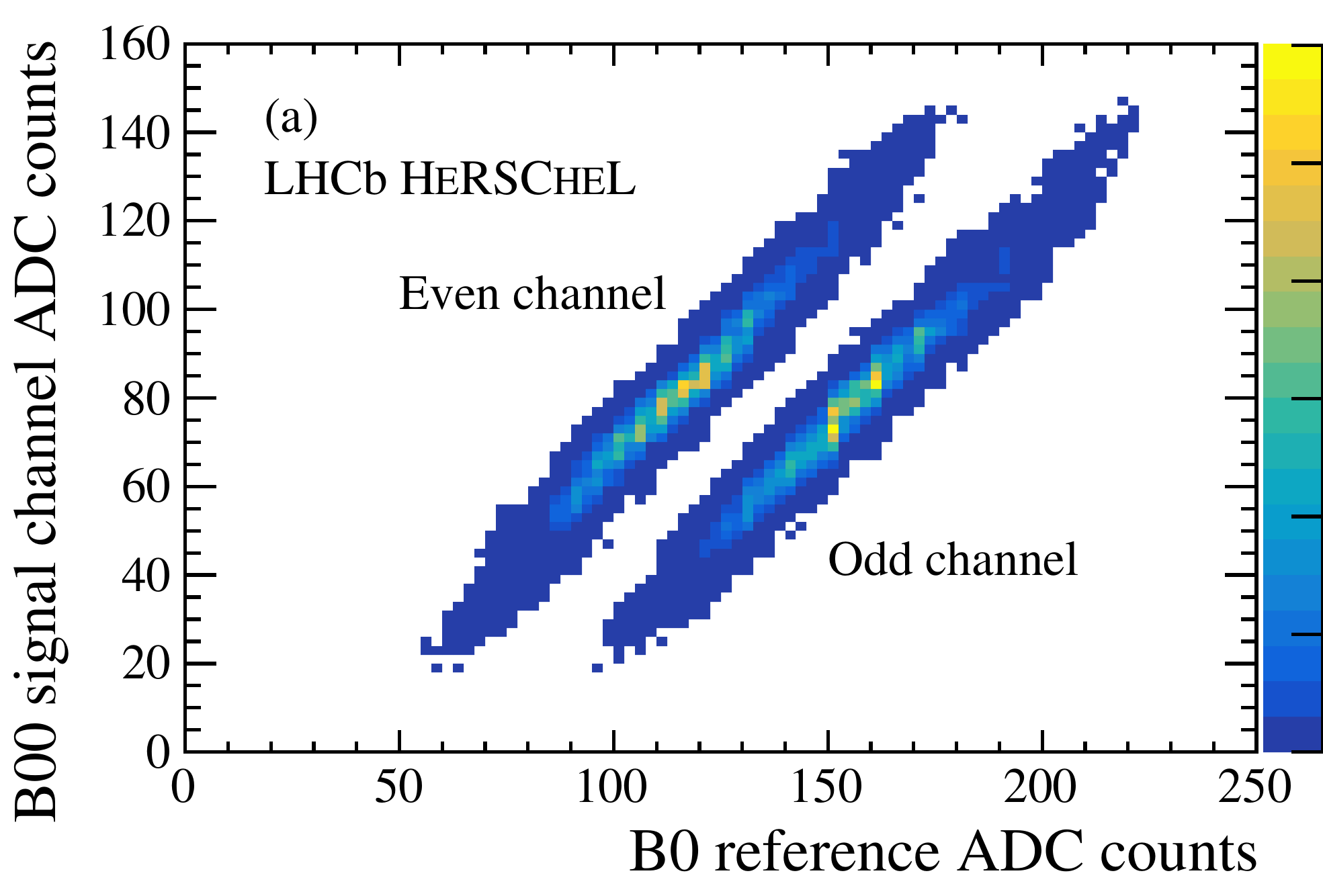}
\includegraphics[width=.49\textwidth]{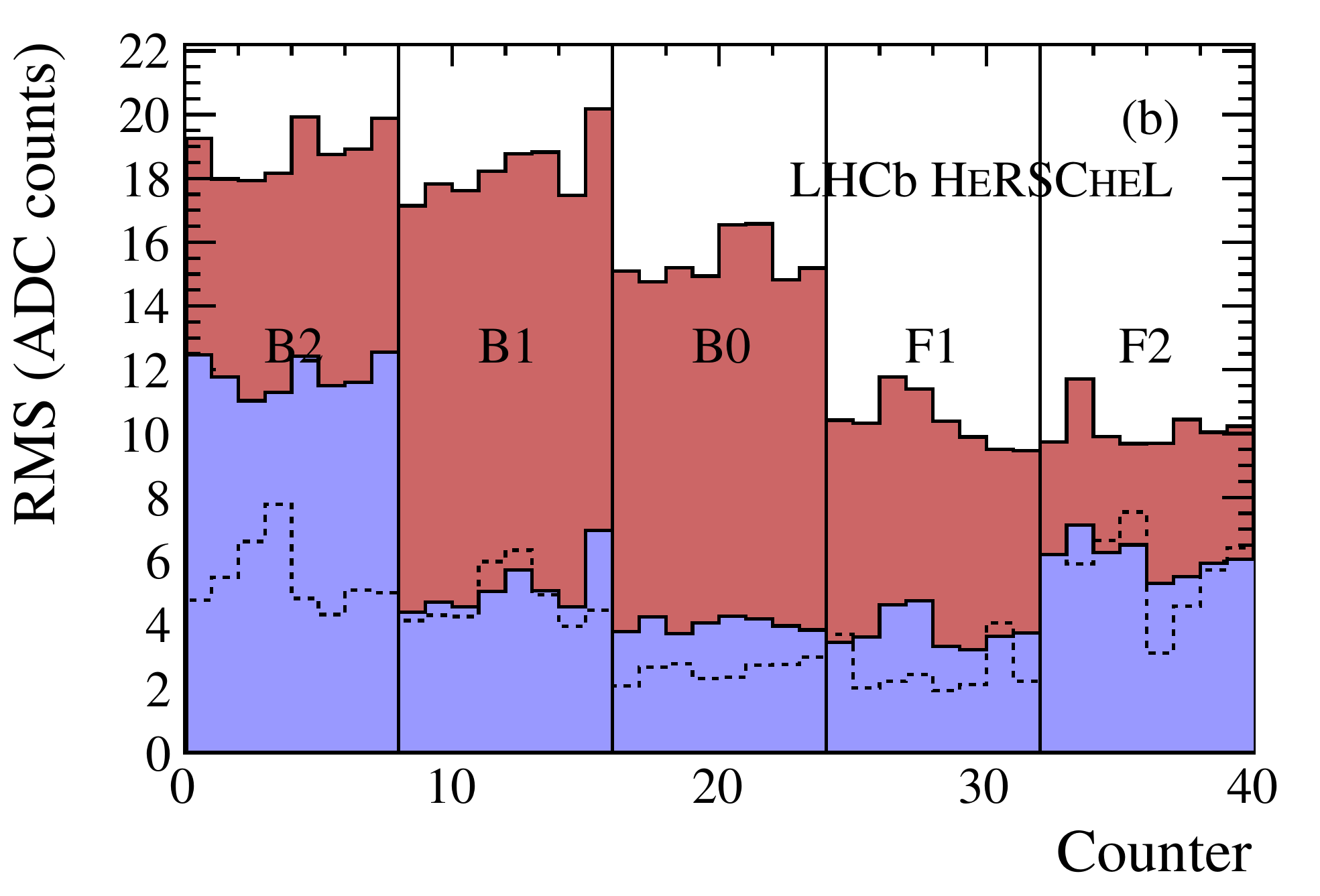}
\caption{Subtraction of common noise. (a) illustrates the correlated noise for the two integrator channels reading out one detector quadrant on station B0 in 2015 end-of-fill data, with respect to the signal in the spare cable (B0 reference), not connected at the detector end. (b) compares the raw signal in end-of-fill data taken during 2015 (red/darker filled histograms) with that after common-noise subtraction (blue filled histogram) and the raw signal  RMS with the new adapter board in 2016 (dashed histogram). Each station houses four PMTs, read out via the dual-channel VFE board, and there are therefore eight channels to consider for each station. The relatively large contribution of uncorrelated noise to the signal in the B2 station was the result of imperfect grounding; this was resolved at the end of 2015.
\label{im:Calibration}}
\end{center}
\end{figure}

The correlation between the signal and spare cables has been studied using data collected during special calibration runs at the end of each fill, once the beams have been dumped. An example of the observed correlation is shown for one channel in Fig.~\ref{im:Calibration}. The common noise is then subtracted offline according to the correlations observed in each fill, and the ADC distribution is shifted to be centred at zero.

The adaptor board was replaced at the end of 2015 operation to employ a less noisy input-matching mechanism. Although data taken during 2016 have not been studied in depth yet, the uncorrected pedestal is seen to be much narrower than that observed in 2015, and no offline noise subtraction is required. The RMS for each channel in end-of-fill data is shown in Fig.~\ref{im:Calibration}, comparing the raw 2015 signal with that after the calibration has been applied, as well as demonstrating the much less noisy raw signal from 2016.

\subsection{Correlation with other LHCb sub-detectors}
\label{sec:correlationWithLHCb}
\begin{figure}[htbp]
\includegraphics[width=.49\textwidth]{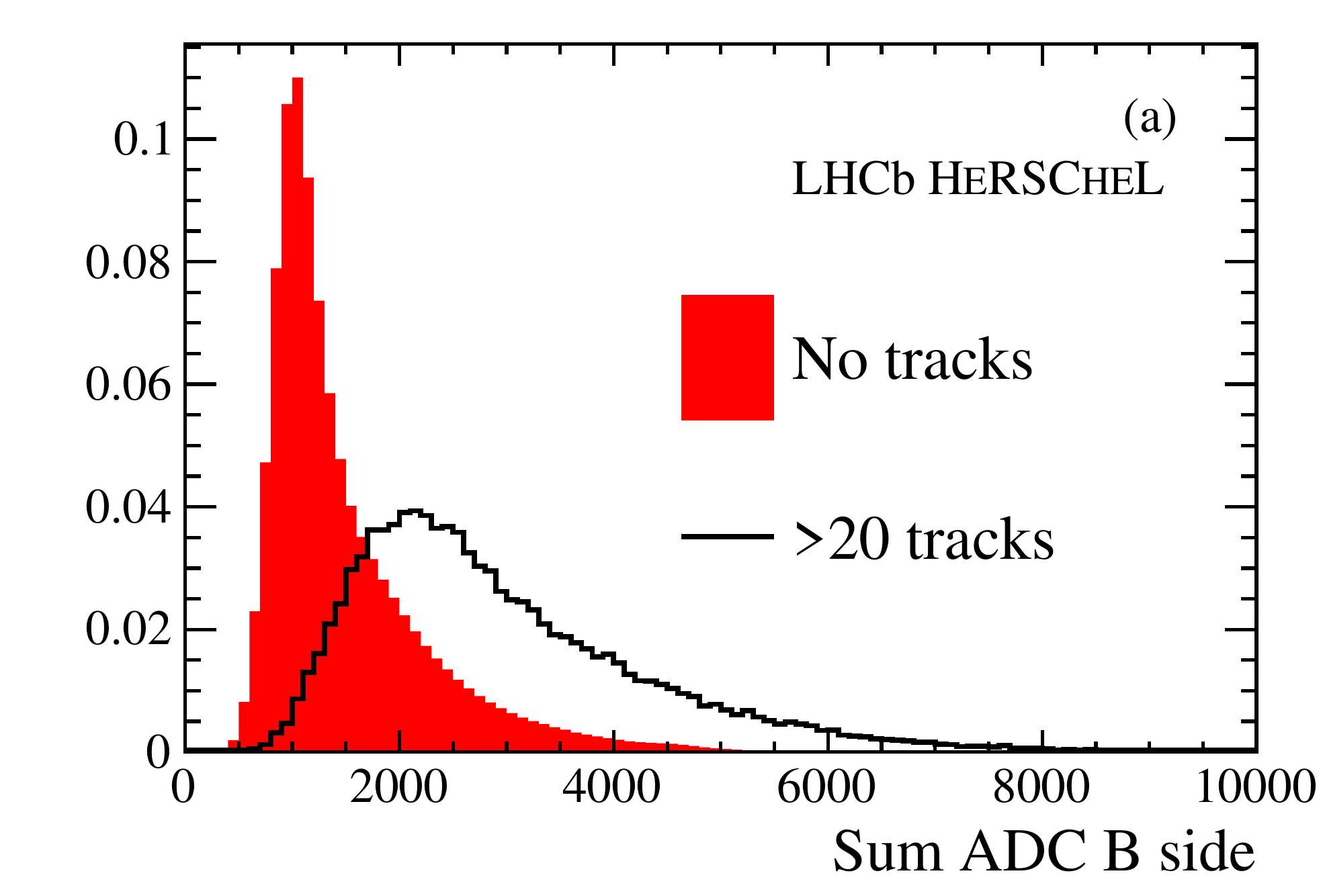}
\includegraphics[width=.49\textwidth]{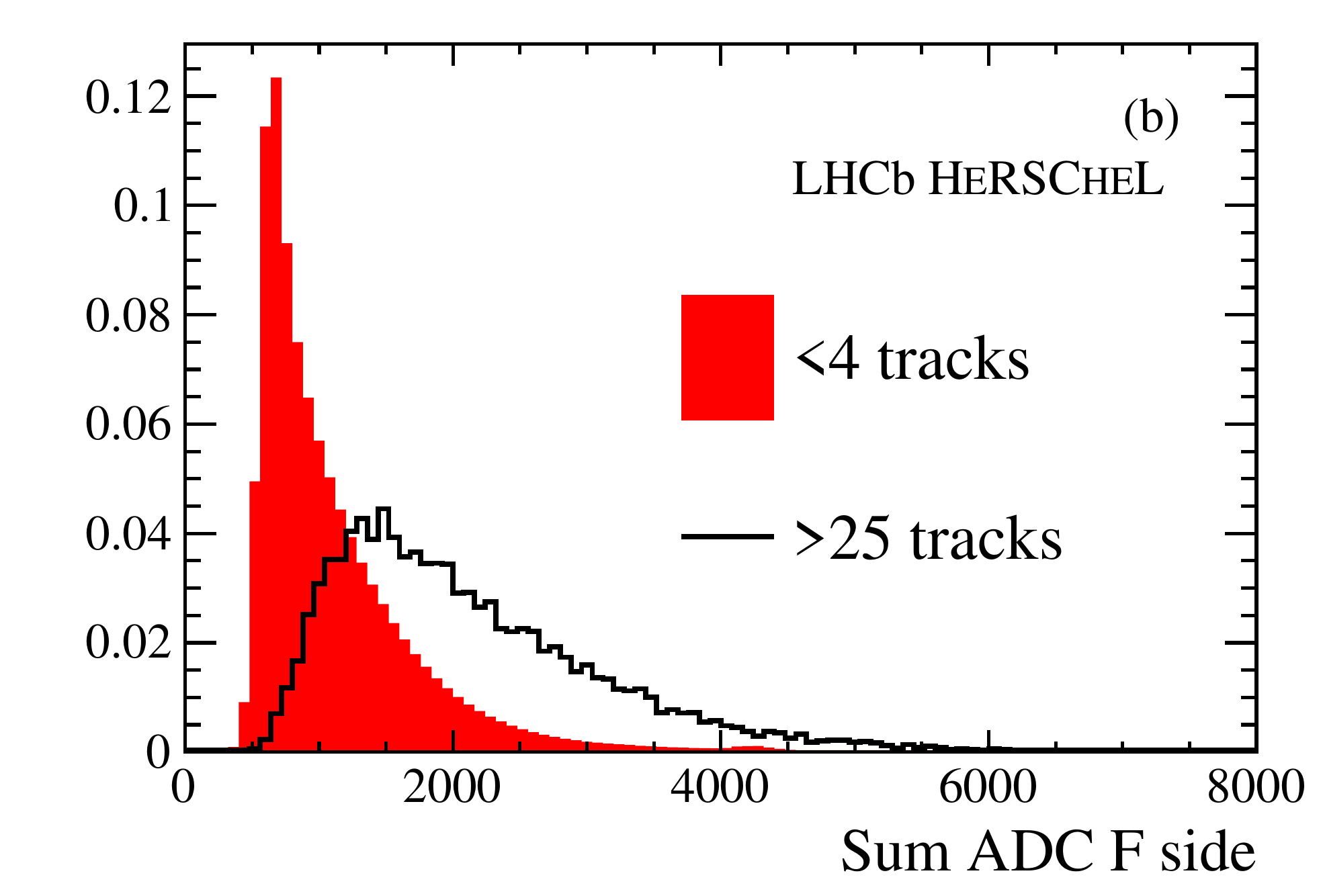}
\caption{Correlation between sum of raw \herschel ADC counts on (a) the B side and (b) the F side of the interaction point, with reconstructed tracks on the same side.} 
\label{im:CorrelationsHerschelVELO}
\end{figure}

An important confirmation of the detector's successful operation is the observation of correspondence between activity registered in other LHCb sub-detectors and that in the \herschel scintillators. In Fig.~\ref{im:CorrelationsHerschelVELO} the sum of the ADC counts from all the stations on each side is shown in the two cases that there are small or large numbers of tracks reconstructed on the corresponding side of the interaction point. For comparison with the {B-side} \herschel response, only `backward' tracks, leaving deposits in the VELO modules behind the interaction point and therefore reconstructed using only that sub-detector, are considered. To compare with the F-side \herschel response, only `long tracks' are employed, where these tracks are reconstructed using deposits in the VELO modules around the interaction point and the tracking stations further downstream.
As expected, more activity is seen in the \herschel counters when more tracks are reconstructed.

\subsection{Empty-detector signal characterisation}
\label{sec:pedestals}
\begin{figure}[htbp]
\includegraphics[width=.32\textwidth]{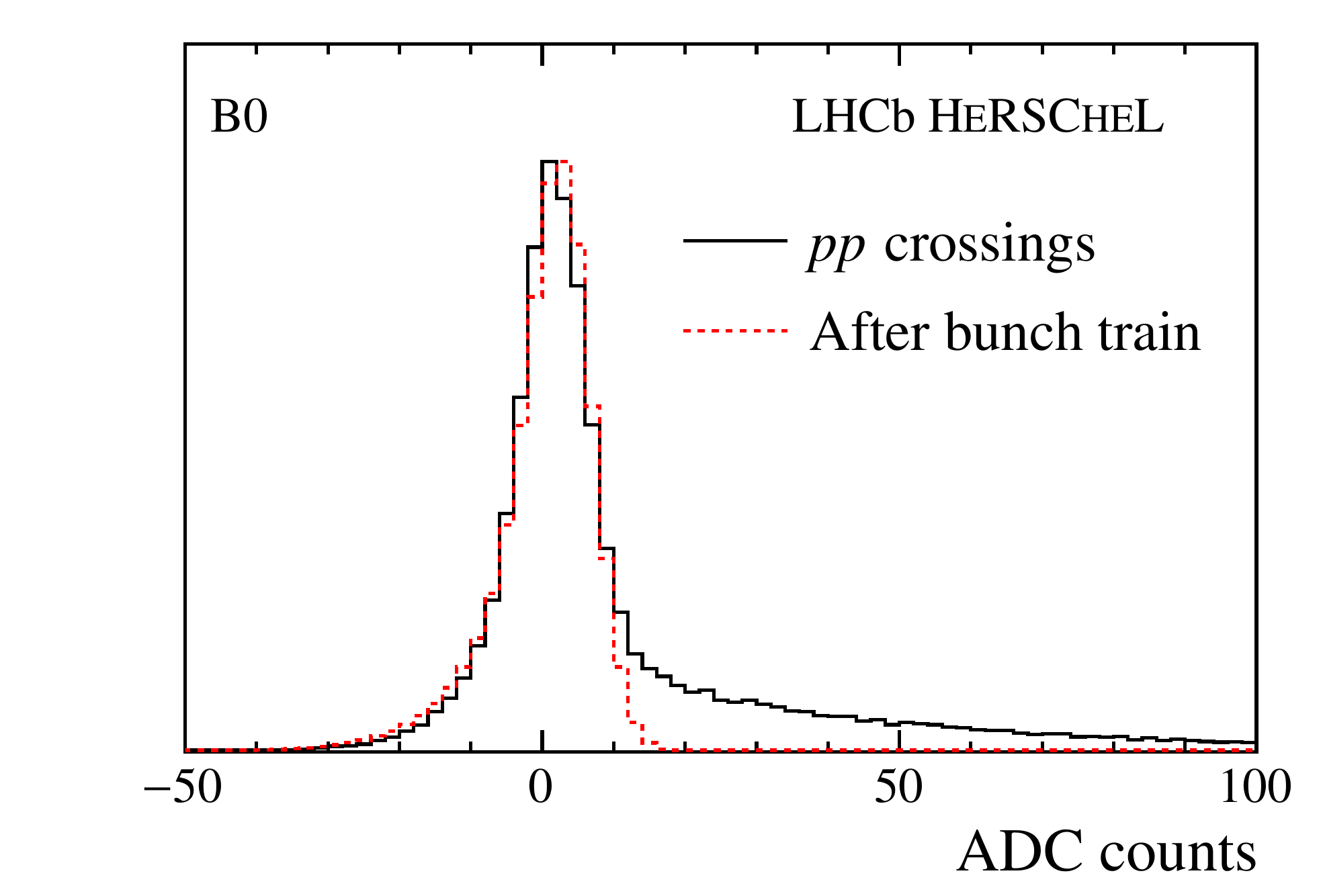}
\includegraphics[width=.32\textwidth]{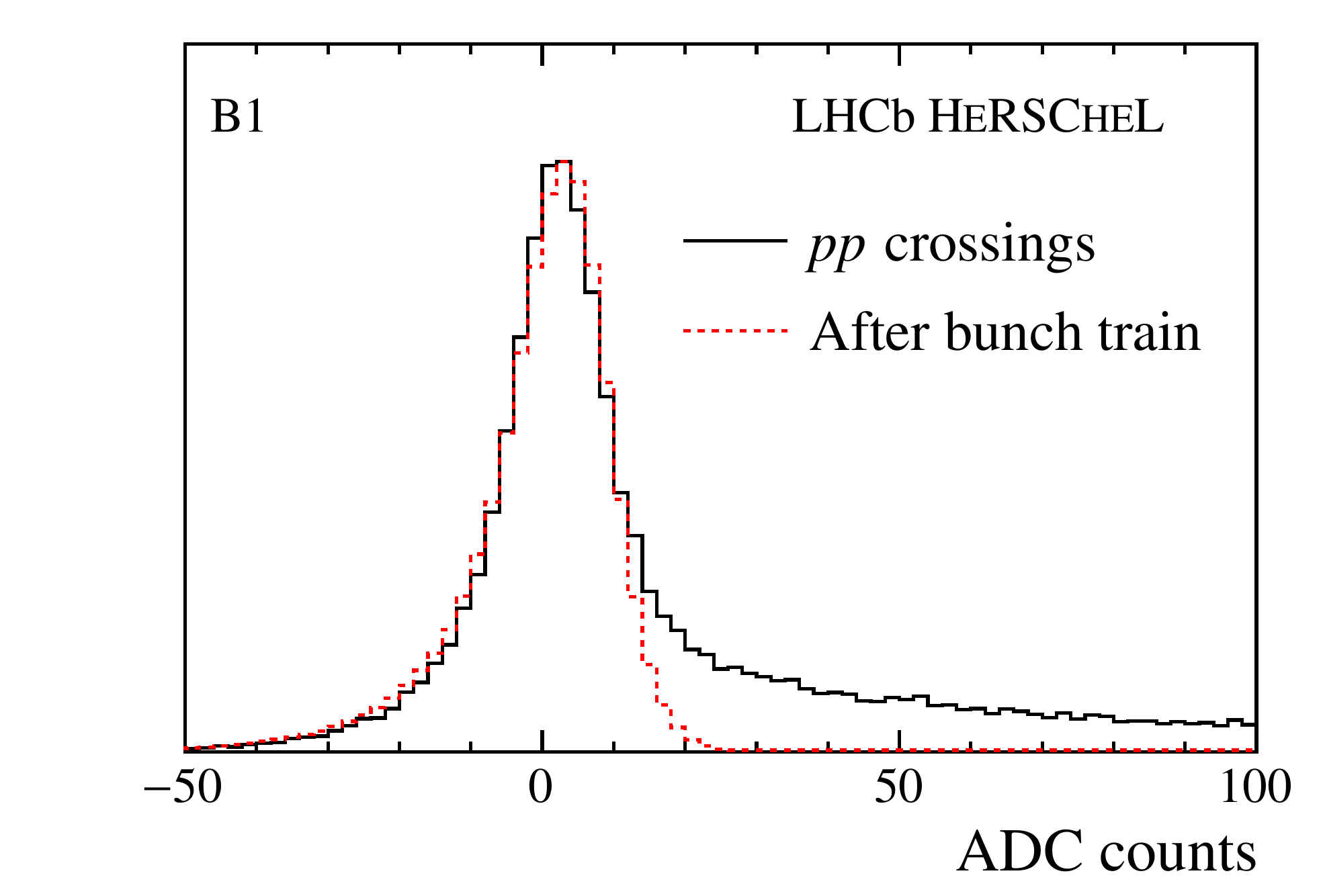}
\includegraphics[width=.32\textwidth]{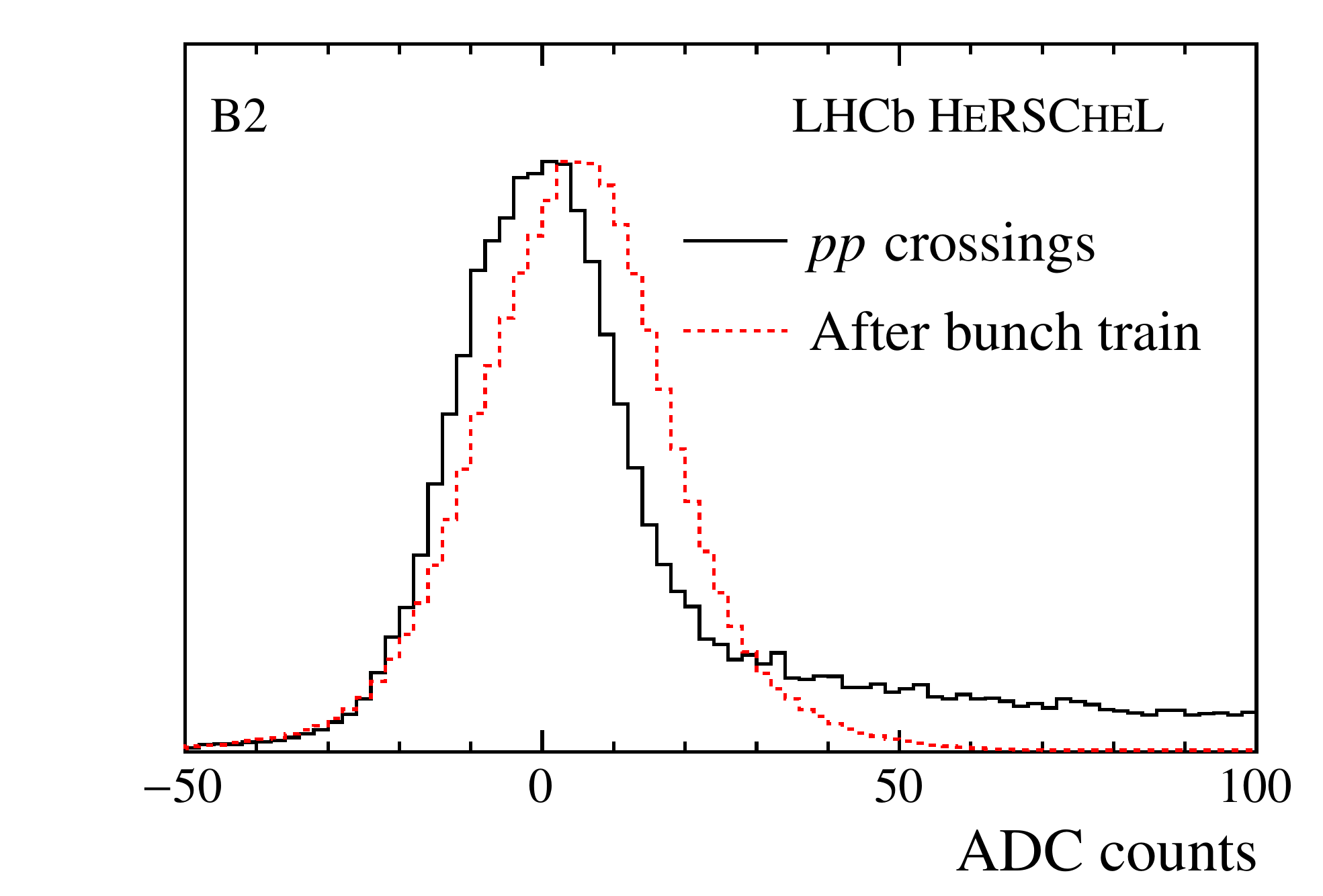}\\
\includegraphics[width=.32\textwidth]{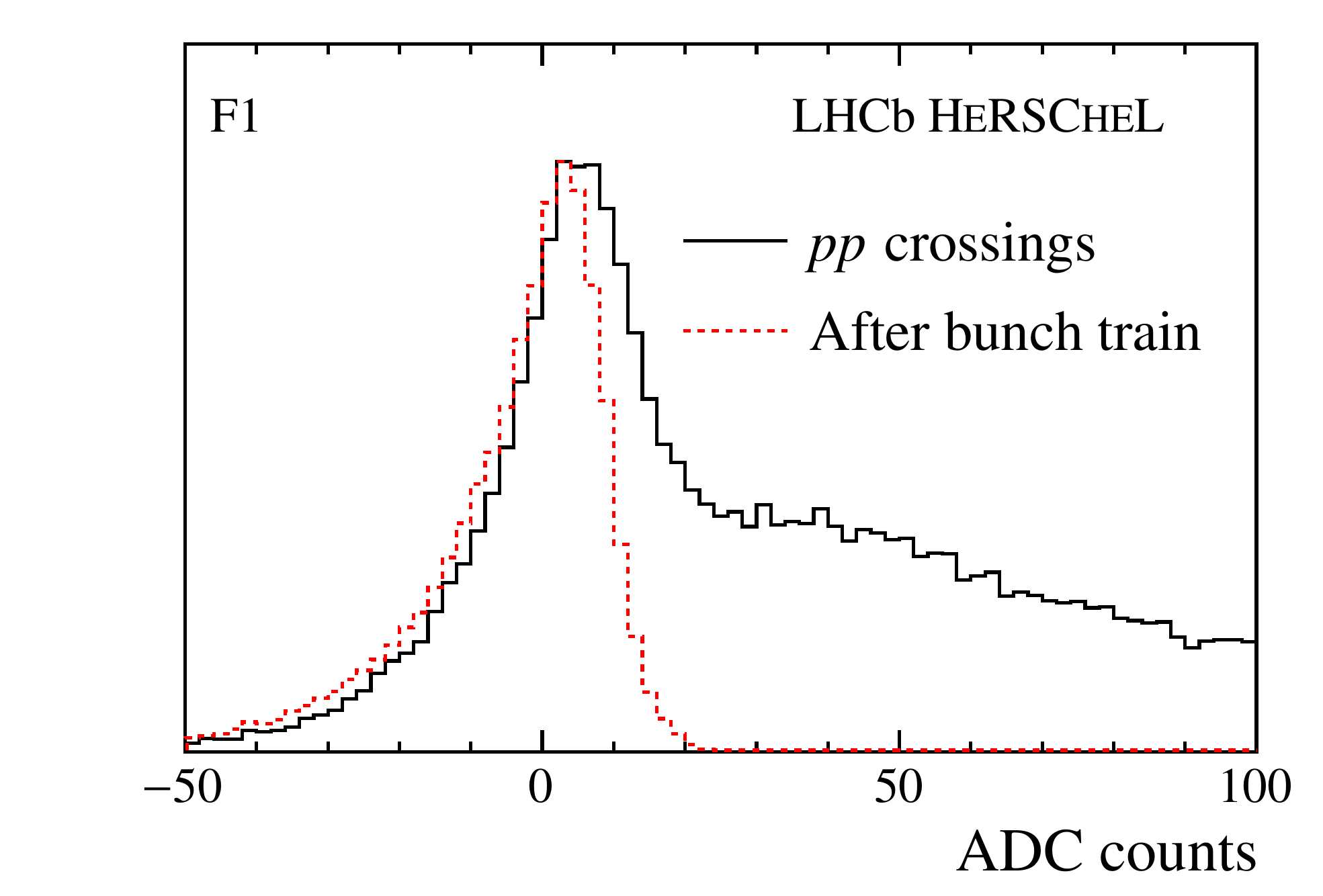}
\includegraphics[width=.32\textwidth]{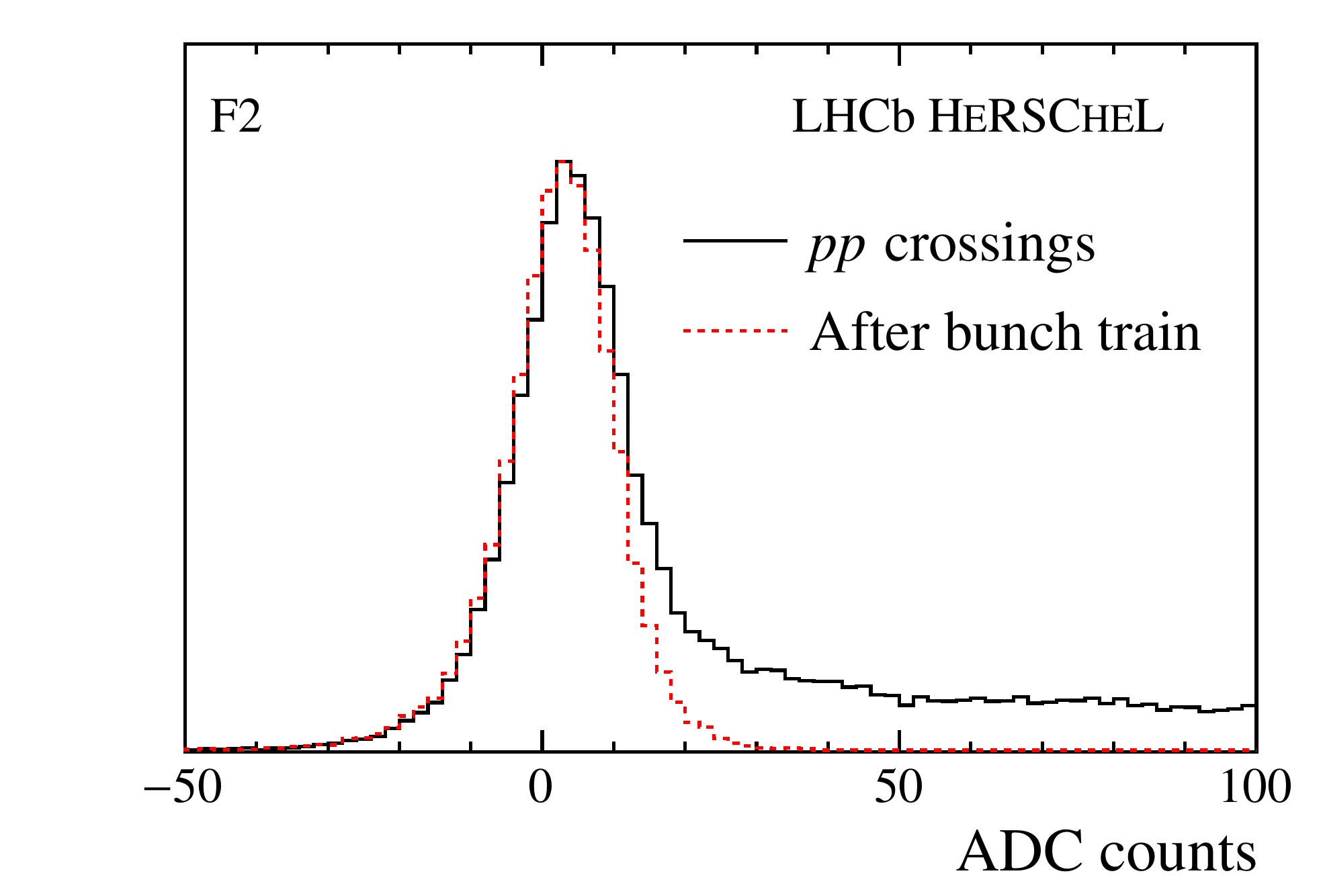}
\caption{Activity registered, after calibration, in one integrator attached to one counter for each \herschel detector station during beam-beam crossings in the solid histogram, showing only the range up to 100 ADC counts. The empty-detector signal recorded after a bunch train is represented by the dotted histogram.\label{im:Pedestals}}
\end{figure}

In this section the expected response of the detector is presented in the case that a single CEP interaction occurs during a $pp$ crossing. For illustration Fig.~\ref{im:Pedestals} shows the signal recorded in one counter at each of the five \herschel detector stations, for an arbitrary sample of $pp$ collision events selected by the LHCb trigger. The two principal features of the distribution are the large portion of events around zero ADC counts in each counter, corresponding to an absence of activity, and a long tail to higher numbers of ADC counts, corresponding to significant activity in the counters. 

Protons circulating in the LHC are distributed in bunches, separated from one another by 25\,\ns. Bunches are collected into `trains' by virtue of the injection procedure, separated by gaps. A 25\,\ns window within which proton bunches cross in LHCb is referred to as a `bunch crossing'. Whilst the dominant contribution to the \herschel empty-detector signal is electronic noise, secondary contributions arise as the result of activity in nearby crossings which spill into the 25\,\ns time interval of the triggered crossing. The largest of these secondary contributions is the residual impact on detector electronics of successive large signals in the detector, during a train of proton-proton crossings in the LHC. It is found that the signal recorded in the counters in the window immediately after such a train, where no particle activity can be present, provides a good description of the empty-detector region of the ADC response, as in the case of a CEP interaction. This ADC response for each of the example counters is also shown in Fig.~\ref{im:Pedestals}.

\section{Impact on physics studies}
\label{sec:physics}

In the case of a single CEP interaction in a $pp$ crossing, no activity is expected in the \herschel detector other than that described in Sect.~\ref{sec:pedestals}. In this section a metric is presented, employing information from all counters in the \herschel detector, that can be used to discriminate between the empty-detector signal and the increased activity associated with background processes. The efficiency of this metric in selecting CEP signal is measured using CEP production of continuum-dimuon pairs, an abundant process that can be rather easily identified without \herschel information. The effectiveness of the same metric in suppressing non-CEP background is determined. Finally the impact of using \herschel information in the measurement of exclusive photoproduction of \jpsi mesons, that can be considered a benchmark for other CEP processes, is considered.

\subsection{Efficiency of \herschel in selecting empty events}
In order to construct a quantity that combines the responses of all twenty counters comprising the \herschel detector effectively, it is beneficial to account for the characteristic distribution of the empty-detector signal in each counter. Examples of this empty detector signal have been discussed in Sec.~\ref{sec:pedestals}. The most natural way to combine the activity in all the \herschel detectors is to construct a $\chi^2$ quantity, $\xi_{\rm HRC}$, such that values of $\xi_{\rm HRC}$ close to zero correspond to events with little or no activity in all the \herschel counters, as expected in the case of a single CEP interaction, and high values of $\xi_{\rm HRC}$ correspond to events where the counter activity is elevated, as expected for non-CEP background. In order to construct such a $\chi^2$ it is necessary to account for two effects. Firstly the non-Gaussian distribution of the empty detector \herschel signals, as displayed in Fig.~\ref{im:Pedestals}, must be accounted for. Secondly it is required to consider the correlation in the activities recorded in the different quadrants of the \herschel stations, even between the different stations on each side, given the overlap in acceptance shown in Fig.~\ref{Fig:SimEtaMinBias}. Both these effects have been studied in events recorded immediately after a series of $pp$ bunch crossings in the LHC. A single $\chi^2$ quantity, $\xi_{\rm HRC}$, is constructed to quantify the activity above the noise in all the counters, taking account of correlations between the counters. 

\begin{figure}[htbp]
\centering
\includegraphics[width=.7\textwidth]{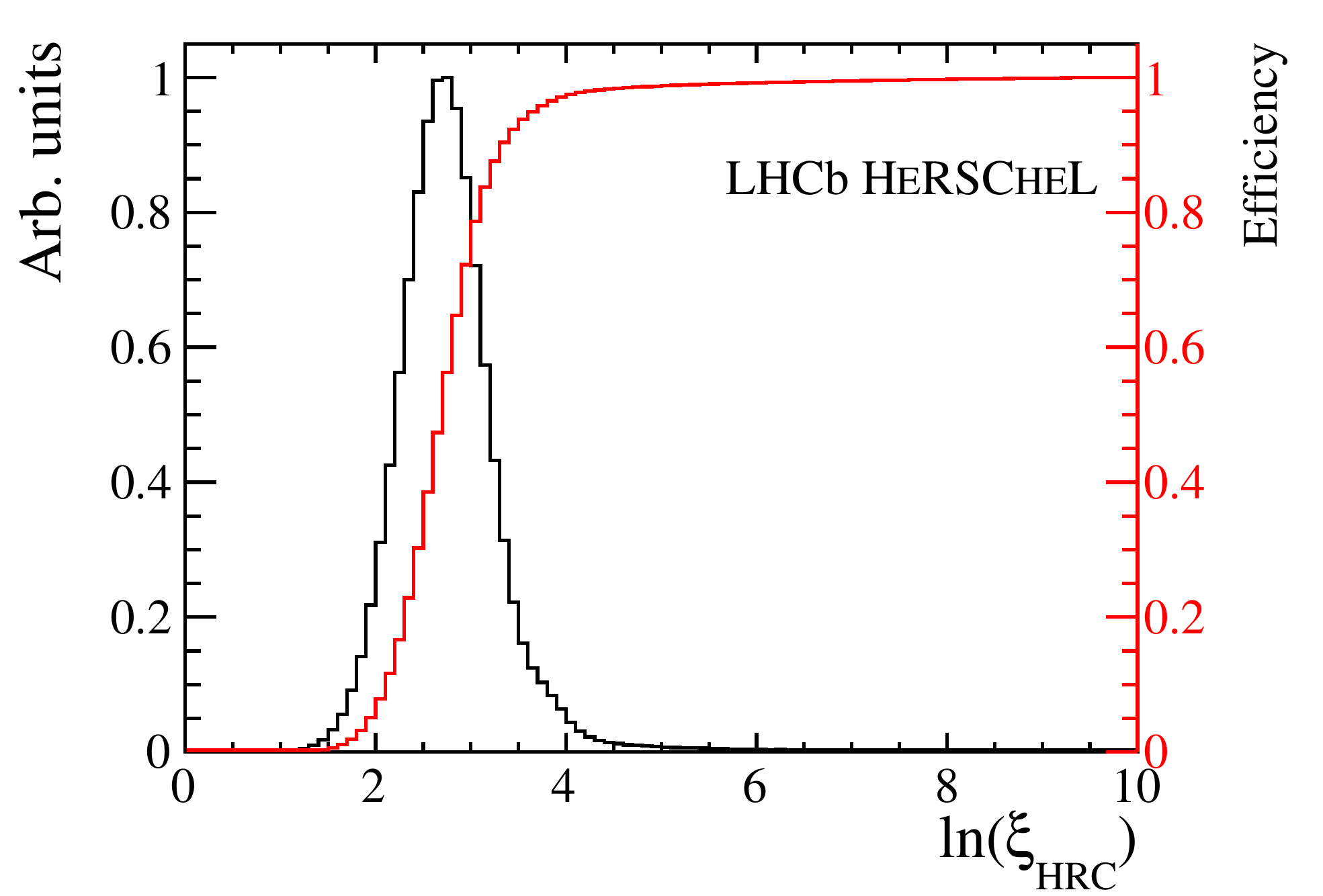}
\caption{Distribution of \herschel metric for events immediately following a bunch train (black line), representative of the response of the detector to CEP signal. The efficiency for an upper limit on the metric for retention of these events is also shown (red line). \label{im:hrcMetricAfterBBtrain}}
\end{figure}

The distribution of the \herschel metric, $\ln(\xi_{\rm HRC})$, is shown in Fig.~\ref{im:hrcMetricAfterBBtrain} for the sample of events immediately after a bunch train discussed in the previous Section. Since these events represent the activity expected in \herschel for CEP processes, the efficiency for these events to survive an upper limit on $\ln(\xi_{\rm HRC})$ is also indicated.

\subsection{Study of the performance on continuum-dimuon production}
\label{sec:sigeff}
In order to enhance the CEP candidates in an event sample, an upper limit on $\xi_{\rm HRC}$ should be chosen. Continuum-dimuon production, an abundant CEP process with a comparatively clear experimental signature, is used to assess the efficiency of such a limit. This process takes place by two-photon exchange between the protons, and results in a very soft spectrum in \ptsq for the dimuon system. By analysing the \ptsq spectrum of the dimuon candidates the CEP component may be reliably isolated.

\begin{figure}[htbp]
\centering
\includegraphics[width=.49\textwidth]{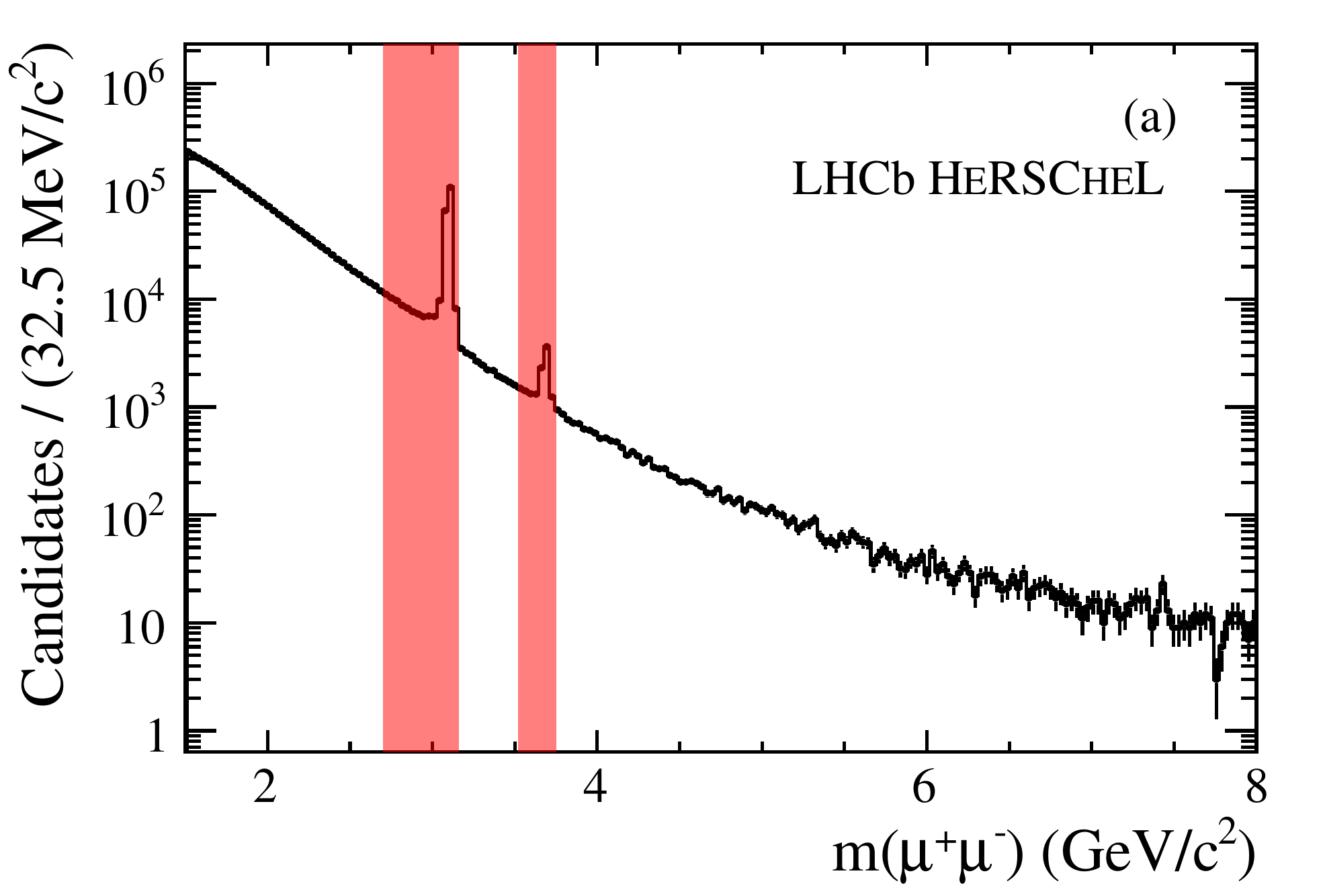}
\includegraphics[width=.49\textwidth]{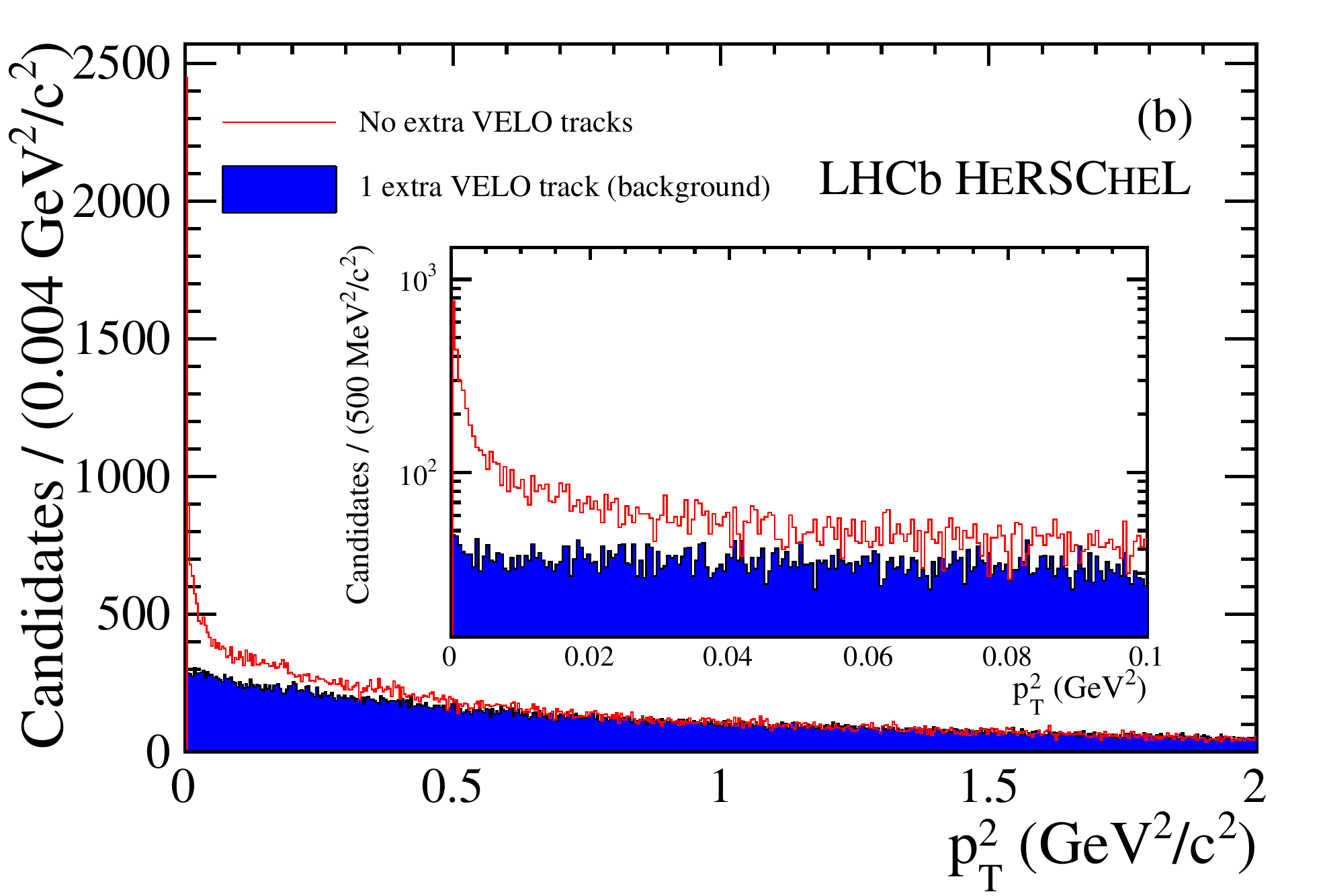}
\caption{Selected continuum-dimuon sample. In (a) is shown the invariant mass distribution of dimuon candidates. The shaded regions of resonant production are excluded. In (b) is shown the full \ptsq distribution of the selected dimuon candidates. 
Since the signal is expected to be concentrated chiefly below 0.1\,\gevgevcc, the candidates below this threshold are shown in the inset histogram. For both regions, an additional histogram is superimposed, corresponding to the distribution of candidates in a background sample.\label{im:DimuonInvMass}}
\end{figure}

A sample of dimuon candidates is selected in data collected during 2015. The fraction of single-interaction CEP candidates within this sample, originating from an event where only a single $pp$ interaction is detected, is enhanced by making standard trigger requirements and imposing limits on the activity in the other LHCb sub-detectors. Specifically only events with exactly two tracks, satisfying standard LHCb muon-identification requirements, are allowed in this two-track sample.
The invariant mass of the candidate, shown in Fig.~\ref{im:DimuonInvMass}(a), is chosen to exclude regions of resonant production, which proceeds by a different CEP mechanism and does not have the same \ptsq distribution. 
The \ptsq distribution of the sample is shown in Fig.~\ref{im:DimuonInvMass}(b). On Fig.~\ref{im:DimuonInvMass}(b) a second distribution has been superimposed corresponding to the \ptsq distribution for dimuon candidates selected in the same way, but where one additional VELO track is required. This category of events is unlikely to be CEP and thus in the case that an additional VELO track is detected, this provides a representation of the background distribution. This distribution is normalised such that the number of background candidates above 0.75\gevgevcc matches the number for the same range in the two-track sample. The very low \ptsq region, where continuum-dimuon CEP is expected to be concentrated, is magnified in the inset histogram. 

\begin{figure}[htbp]
\centering
\includegraphics[width=.49\textwidth]{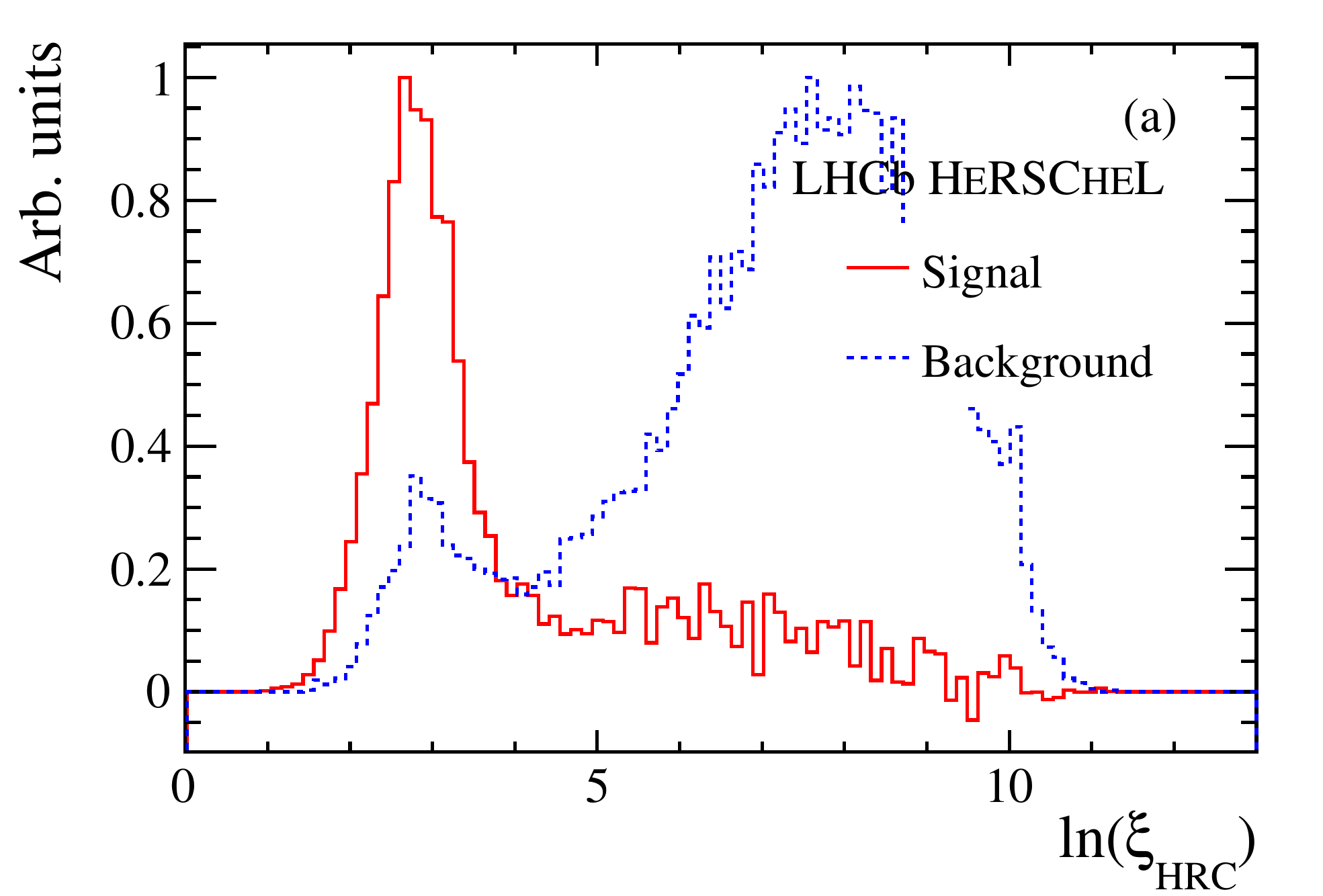}
\includegraphics[width=.49\textwidth]{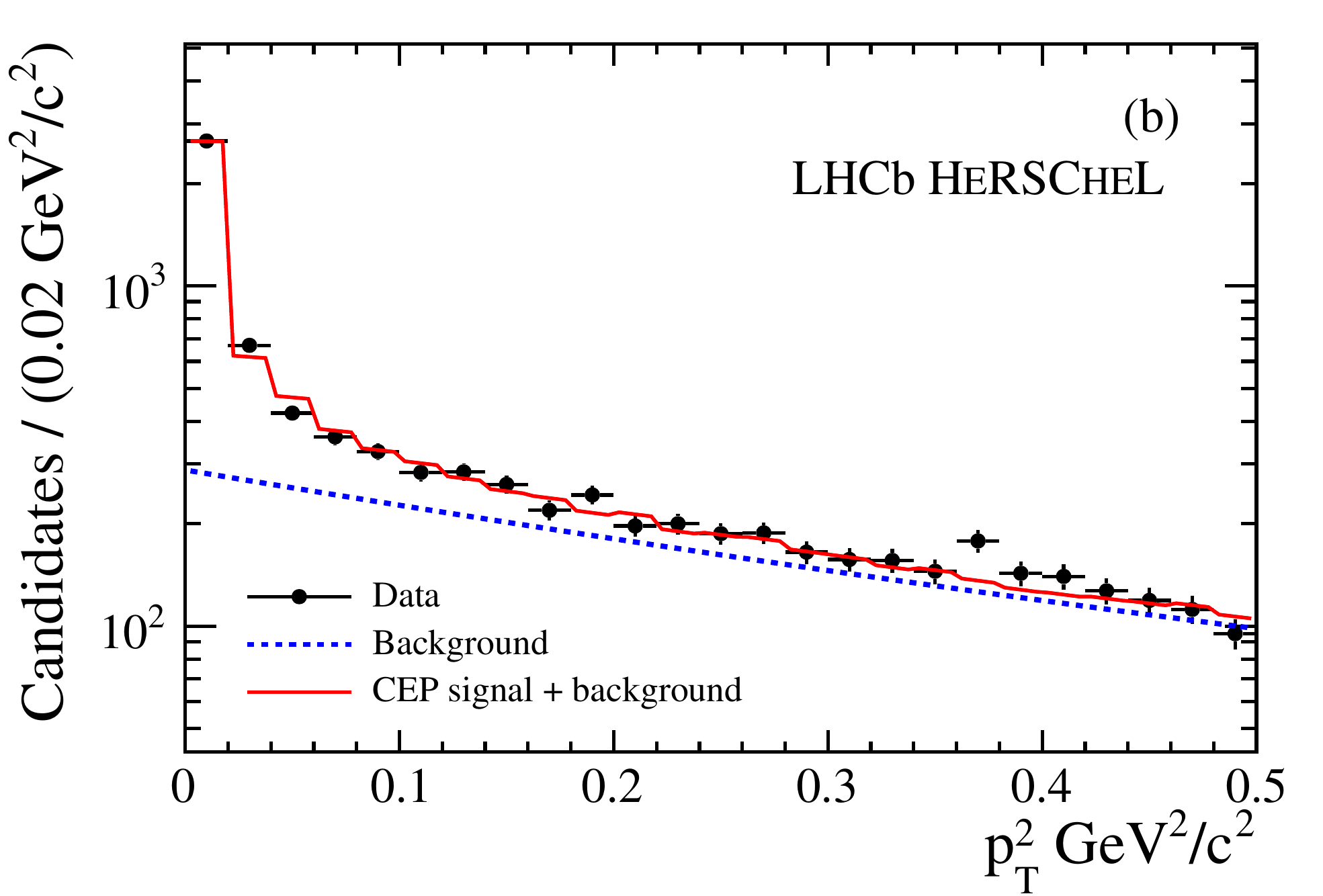}\\
\includegraphics[width=.49\textwidth]{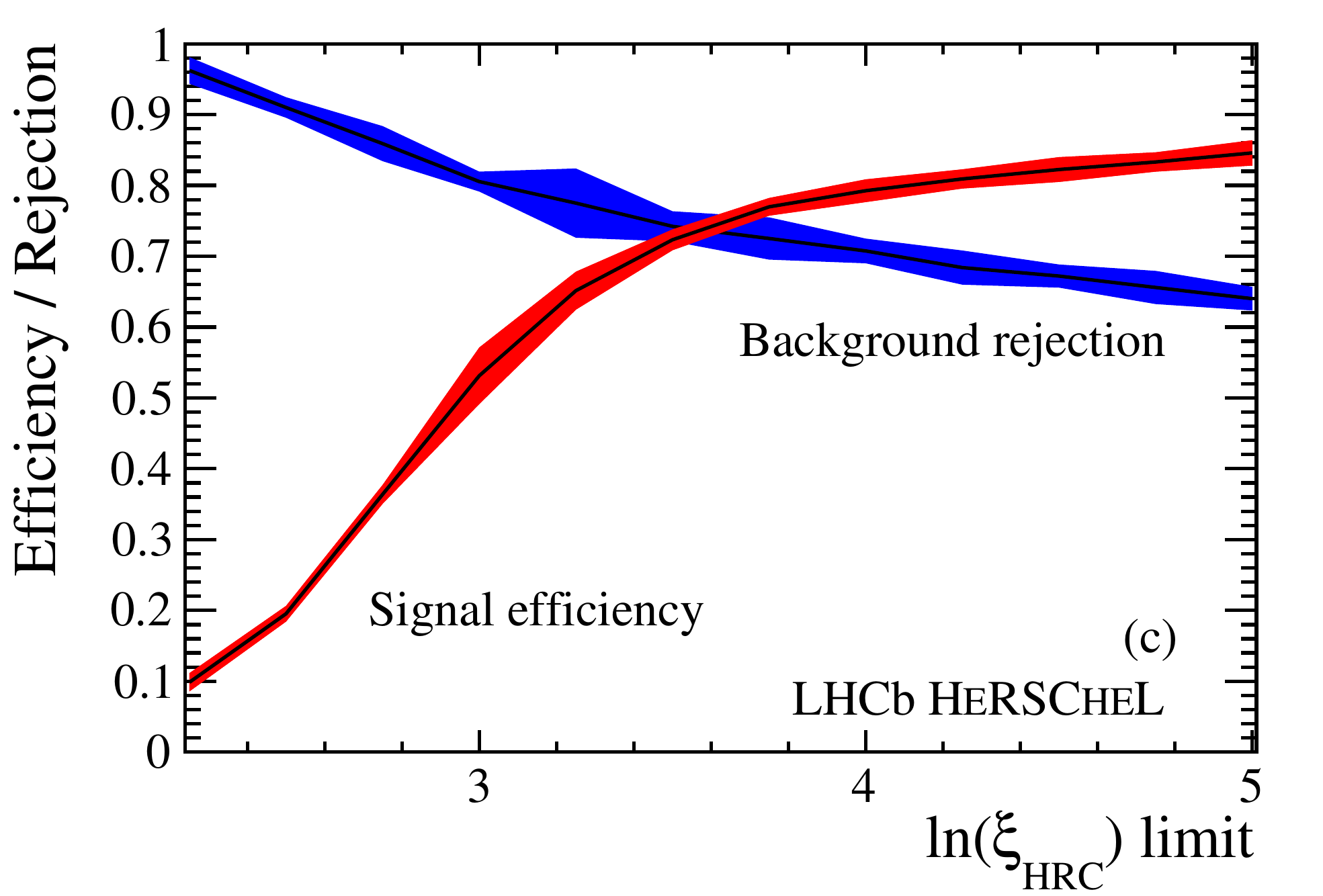}
\caption{In (a) the distribution of $\ln(\xi_{\rm HRC})$ is shown for CEP (solid line) and non-CEP (dashed line) continuum-dimuon candidates. In (b) an example fit to the continuum-dimuon \ptsq distribution is shown, with the CEP signal template and double-exponential background indicated. In (c) the CEP signal efficiency and, for this process, background rejection are shown as a function of the limit chosen on $\ln(\xi_{\rm HRC})$.\label{im:DimuonCEPChi2}}
\end{figure}

These samples are used to obtain the response of the \herschel metric to exclusive continuum-dimuon production. The distribution of the natural logarithm of $\xi_{\rm HRC}$ in the background sample is represented by the dashed line in Fig.~\ref{im:DimuonCEPChi2}(a). To visualise the $\ln(\xi_{\rm HRC})$ response for CEP signal, the distribution is first obtained for dimuon candidates with \ptsq below 0.1\gevgevcc in the signal sample. The background shape can be subtracted from this distribution according to the number of events in the background sample, again normalised in the high-\ptsq region. The signal response is shown as the solid line in Fig.~\ref{im:DimuonCEPChi2}(a). It is interesting to note that the background sample has a peak at low values of $\ln(\xi_{\rm HRC})$, a feature which can be explained by the presence of background where all additional activity associated with, for example,  proton dissociation lies outside the \herschel acceptance. The significant tail in \herschel activity seen even for `CEP signal' appears because, although the activity in the LHCb spectrometer is consistent with only one visible $pp$ interaction there, a second pile-up interaction may still occur producing activity only in the geometrical acceptance covered by the \herschel detector.

In order to obtain the signal efficiency of a given limit on activity in the \herschel detector, the continuum-dimuon sample can again be used. This time a fit is performed to the \ptsq distribution of signal candidates. A template is taken from simulated samples to represent the signal component. Studies of the three-track background sample and other, associated, data sets give confidence that the background component can be well-modelled using a probability density function (PDF) consisting of two exponential shapes. To obtain the efficiency of the \herschel activity limit on the continuum-dimuon CEP signal, the \ptsq distribution is fit with a PDF containing the signal template and the background model, where the slope parameters of the background exponentials are free to vary along with their relative fraction. One such fit is shown, for illustration, in Fig.~\ref{im:DimuonCEPChi2}(b). The effect on the exclusive-signal yield of changing the \herschel veto is determined and shown  in Fig.~\ref{im:DimuonCEPChi2}(c), where the uncertainty is systematically dominated.

It is of interest to consider the effect of the \herschel activity limit on the background component, although it should be noted that the kinematic properties and likely \herschel response to the background are rather specific to the production mode under consideration. The effect of the \herschel activity limit in suppressing the background in this specific sample is also shown in Fig.~\ref{im:DimuonCEPChi2}(c), where the uncertainty receives comparable contributions from statistical and systematic sources.

An illustrative veto on activity in \herschel is chosen, consisting of an upper limit on $\ln(\xi_{\rm HRC})$ at 4.9. The signal efficiency for continuum-dimuon production at this working point is 84\% according to Fig.~\ref{im:DimuonCEPChi2}(c) and, as seen in the same study, 65\% of the background is rejected. It is clear from Fig.~\ref{im:hrcMetricAfterBBtrain} that the intrinsic \herschel efficiency for selection of empty events at this working point is nearly 100\%. The difference between the empty-event selection efficiency and the measured efficiency for selection of CEP continuum-dimuon candidates arises because a limit on activity in the \herschel detector also suppresses events where additional, pile-up $pp$ interactions take place. Such additional pile-up $pp$ interactions, if the resulting particle production escapes the standard spectrometer acceptance but enters the \herschel detector, would be likely to produce a background-like signature in some fraction of events also containing a genuine CEP candidate.

\subsection{Study of the performance on \jpsi photoproduction}
\begin{figure}[htbp]
\centering
\includegraphics[width=.49\textwidth]{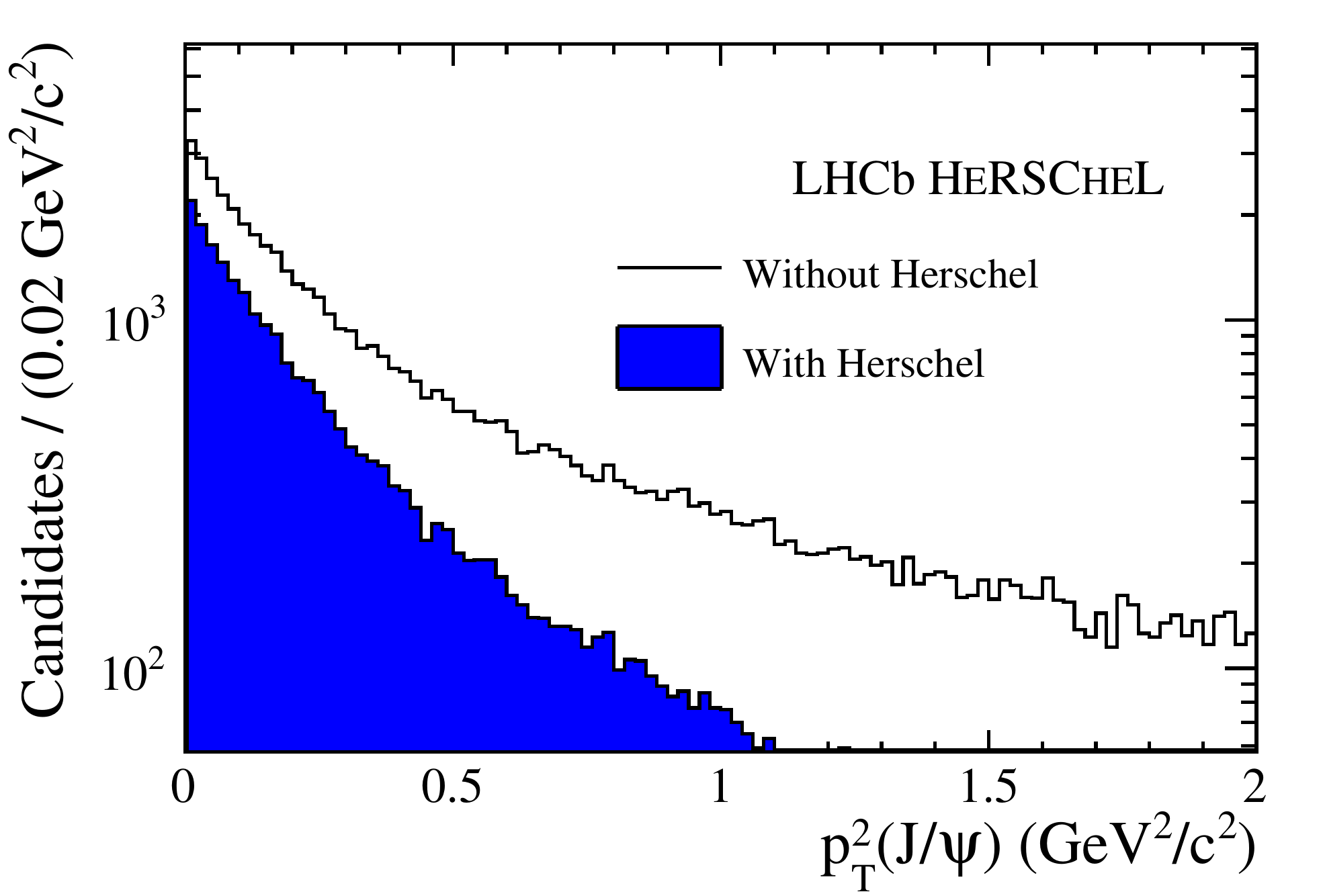}
\caption{Effect on the \ptsq distribution of \jpsi CEP candidates of applying a limit of 4.9 on $\ln(\xi_{\rm HRC})$.  This cut is expected to retain close to 100\% of CEP signal which have no additional tracks due to pile-up or single diffractive events in the combined LHCb and \herschel geometrical acceptance. This corresponds to about 84\% of the events which would be accepted if only the LHCb geometrical acceptance was used. At this working point, the yield of non-CEP \jpsi candidates falls to nearly a third of the original value.\label{im:jpsiCEPptsq}}
\end{figure}

The effect of the limit placed on $\xi_{\rm HRC}$ can be considered for the case of resonant dimuon production, near the \jpsi mass indicated by the first shaded region in Fig.~\ref{im:DimuonInvMass}(a). In this case, the production mechanism is predominantly photoproduction, neglecting a very small contamination from continuum-dimuon production in this region, and the exclusive \jpsi candidate \ptsq distribution extends to higher values, even up to $1\gevgevcc$. The effect of placing a limit on $\ln(\xi_{\rm HRC})$ of 4.9, chosen to have a signal efficiency of approximately 84\% according to the continuum-dimuon studies, is shown for the distribution of \jpsi candidate \ptsq in Fig.~\ref{im:jpsiCEPptsq}. The background yield, determined by the number of candidates in the high \ptsq region above 1\,\gevgevcc, where little \jpsi CEP signal is expected, is reduced to less than a third of its original value. A similar suppression is expected in the background component at lower \ptsq, and a corresponding improvement in the signal purity. The background rejection that is achieved at a particular \herschel working point depends on the production process for that background, hence for this process the \herschel cut is more effective in suppressing background than in the continuum-dimuon study. 

\section{Employing \herschel in the LHCb trigger}
\label{sec:hrcInTrigger}
\begin{figure}[htbp]
\centering
\includegraphics[width=.32\textwidth]{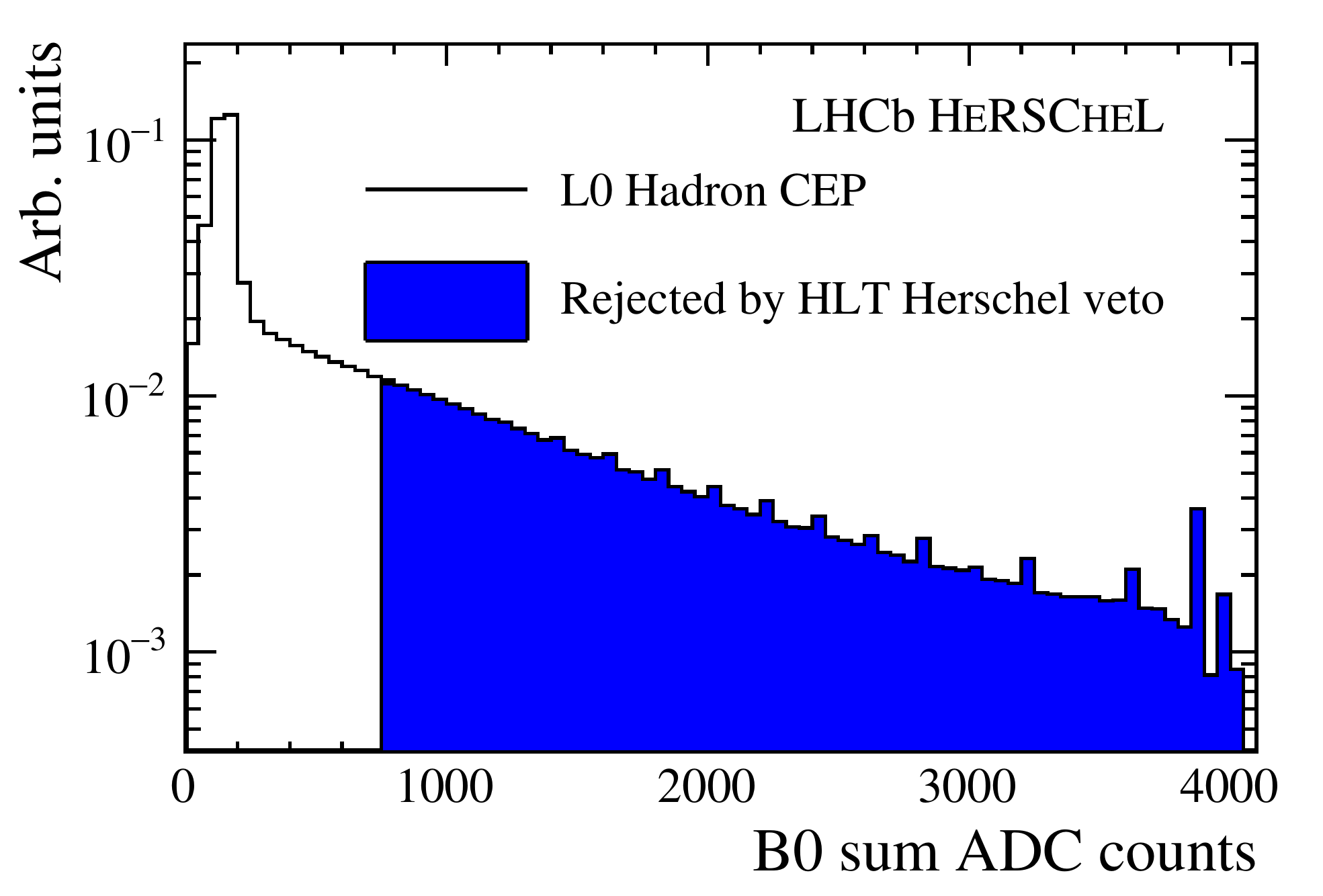}
\includegraphics[width=.32\textwidth]{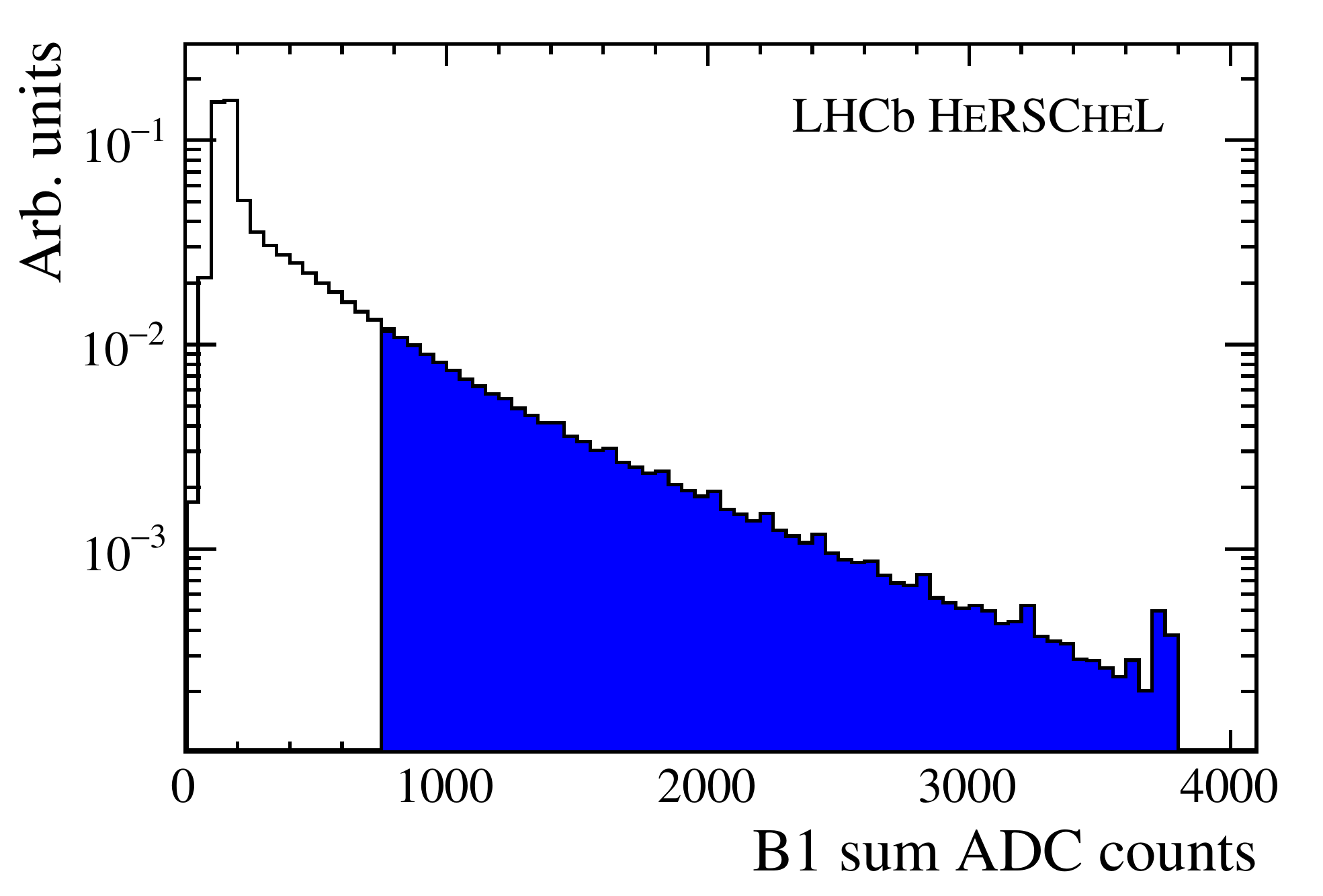}
\includegraphics[width=.32\textwidth]{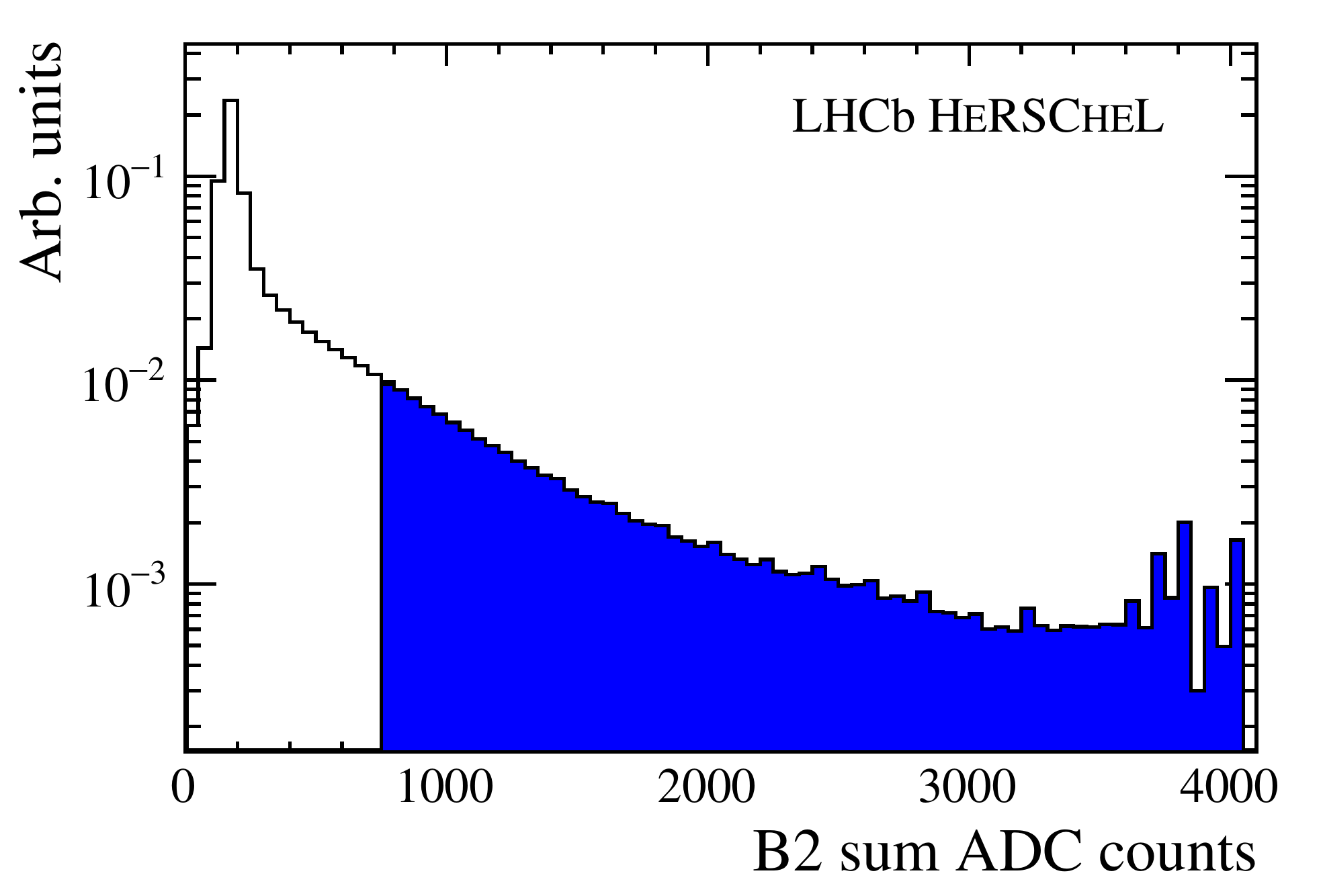}\\
\includegraphics[width=.32\textwidth]{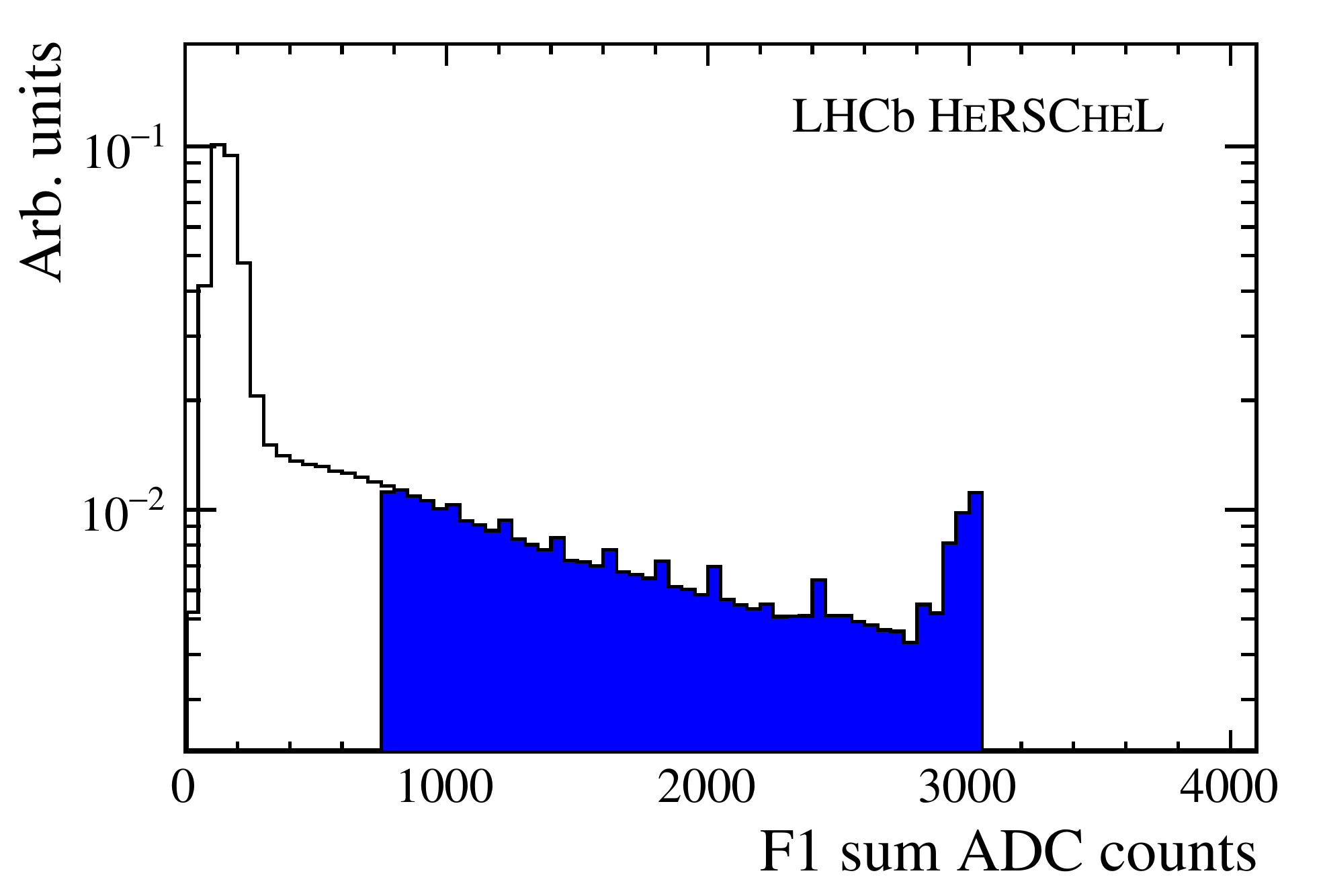}
\includegraphics[width=.32\textwidth]{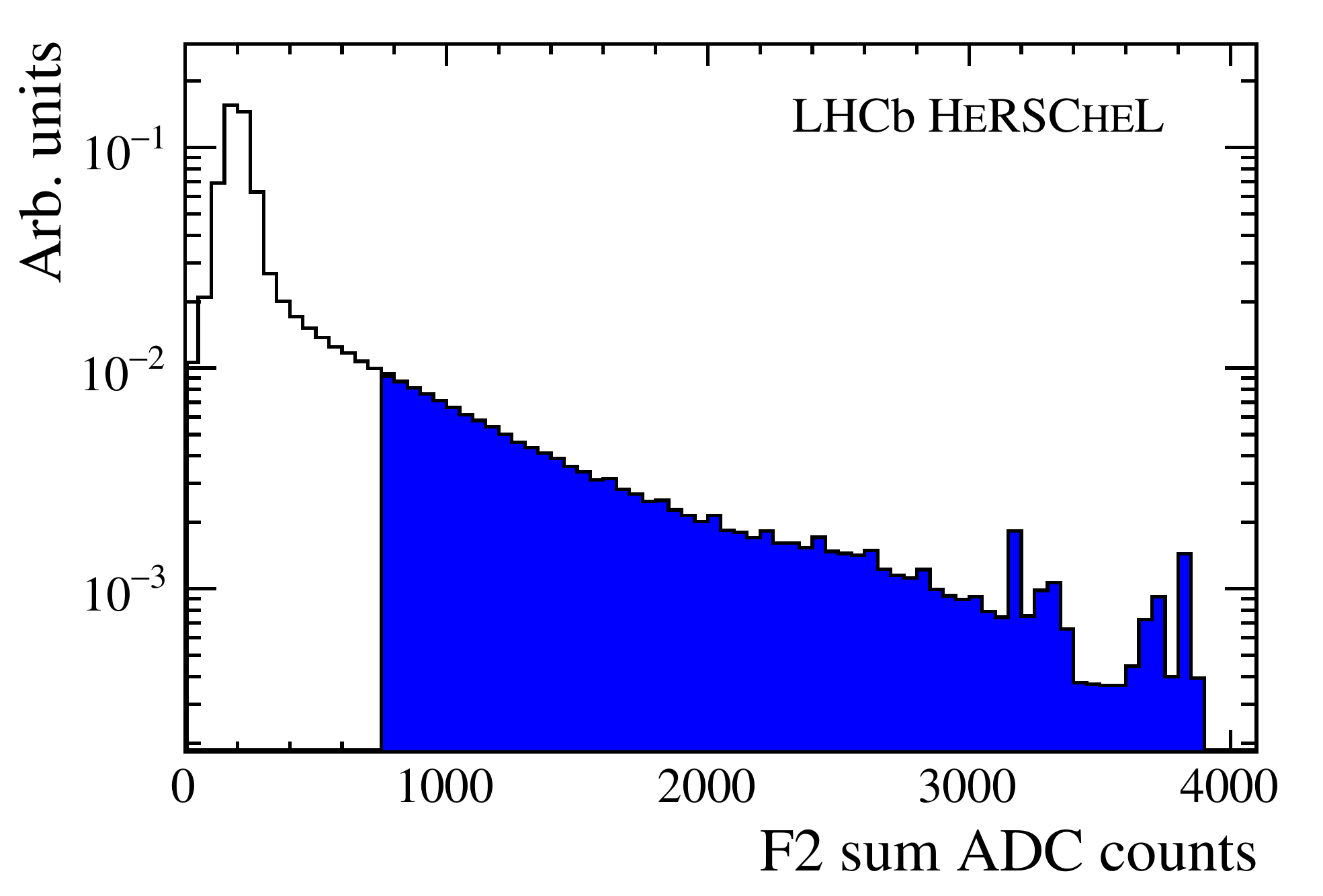}
\caption{\herschel in the LHCb software trigger. Distributions of the sum of ADC counts recorded by the four counters in each station. The portion of events rejected by the limit in each station is indicated by the filled region.\label{im:triggerSums}}
\end{figure}

The LHCb trigger is divided into a low-level hardware trigger and a higher-level software trigger. In 2016 \herschel information was introduced into the software trigger. Specifically, a loose limit is placed on the sum of  ADC counts for the counters in each station. The distributions of the sum of ADC counts from each station are shown in Fig.~\ref{im:triggerSums} for events which pass the hardware trigger relevant for hadronic CEP. The portion of the distribution rejected is indicated. This conservative limit, far above the empty-detector portion of the distribution, allows for the trigger rate to be reduced by nearly a factor of two, allowing for other constraints to be relaxed in order to select hadronic CEP signals more efficiently. Work is ongoing to include \herschel information in the hardware trigger.

\section{Conclusion and outlook}
\label{sec:conclusions}
A new sub-detector has been constructed, installed and commissioned in the LHC tunnel in order to extend the rapidity coverage of the LHCb experiment in the high-rapidity regions either side of the interaction point during Run 2 of the LHC. Five stations have been installed up to a maximum distance of approximately 114\,m from the interaction point. The \herschel detector extends the LHCb forward coverage up to a pseudorapidity of around 10.

The physics impact of the new detector has been described; in particular \herschel allows for significant improvement in the suppression of the most problematic backgrounds in studies of Central Exclusive Production at LHCb, whilst maintaining high signal efficiency. The new detector is already in use in the LHCb software trigger for CEP processes, and will be integrated at the hardware level in the L0 trigger chain in the near future.

\section*{Acknowledgements}
\noindent We thank all our LHCb collaborators who have contributed to the design, construction, commissioning, and operation of \herschel, in particular J. Buytaert, P. Dziurdzia, C. Gaspar, A. Granik, Yu. Guz, C. Joram, F. Machefert, P. Robbe, and B. Schmidt. Furthermore we are grateful to J. Bernhard from the COMPASS collaboration who provided us with the PMT housings. We express our gratitude to our colleagues in the CERN accelerator departments for the excellent performance of the LHC. We thank the technical and administrative staff at the authors' institutes. We acknowledge support from CERN and from the national agencies: CAPES, CNPq, FAPERJ and FINEP (Brazil); CNRS/IN2P3 (France); MNiSW and NCN (Poland); MinES and FASO (Russia); MinECo (Spain); STFC (United Kingdom). We acknowledge the computing resources that are provided by CERN, IN2P3 (France), KIT and DESY (Germany), INFN (Italy), SURF (The Netherlands), PIC (Spain), GridPP (United Kingdom), RRCKI and Yandex LLC (Russia), CSCS (Switzerland), IFIN-HH (Romania), CBPF (Brazil), PL-GRID (Poland) and OSC (USA). We are indebted to the communities behind the multiple open-source software packages on which we depend.

\addcontentsline{toc}{section}{References}
\setboolean{inbibliography}{true}
\bibliographystyle{LHCb}
\bibliography{main,LHCb-PAPER,LHCb-CONF,LHCb-DP,LHCb-TDR}

\end{document}